\documentclass[aps,prb,floatfix,amsmath,amssymb,preprint,eqsecnum,nofootinbib]{revtex4}
\usepackage{graphicx}
\usepackage{dcolumn}
\usepackage{bm}
\usepackage{epstopdf}
\usepackage{color}

\newcommand{\beq}{\begin{equation}}
\newcommand{\eeq}{\end{equation}}
\newcommand{\ba}{\begin{array}{ccc}}
\newcommand{\ea}{\end{array}}
\newcommand{\nn}{\nonumber \\}

\def\bea{\begin{eqnarray}}
\def\eea{\end{eqnarray}}

\usepackage{setspace} 
\usepackage{amsmath}
\usepackage{amssymb}
\begin{document}

\title{The Landscape of the Hubbard Model}

\author{Subir Sachdev}
\affiliation{Department of Physics, Harvard University, Cambridge MA
02138}

\date{\today \\
\vspace{1.6in}}

\begin{abstract}
I present a pedagogical survey of a variety of quantum phases
of the Hubbard model. The honeycomb lattice model has a
conformal field theory connecting the semi-metal to the insulator
with N\'eel order. States with fractionalized excitations are linked
to the deconfined phases of gauge theories. I also consider the confining
phases of such gauge theories, and show how Berry phases of monopoles
induce valence bond solid order. 
The triangular lattice model can display a metal-insulator transition from a
Fermi liquid to a deconfined spin liquid, and I describe the theory of this transition.
The bilayer triangular lattice is used to illustrate another
compressible metallic phase, the `fractionalized Fermi liquid'.
I make numerous connections of these phases and critical points to the AdS/CFT correspondence.
In particular, I argue that two recent holographic constructions connect respectively to the Fermi liquid
and fractionalized Fermi liquid phases.
\\
~\\
~\\
{\tt TASI Lectures (Boulder, June 2010)\\ ~\\
Chandrasekhar Lecture 
Series and Discussion Meeting on ``Strongly Correlated Systems and AdS/CFT''  (International Center for Theoretical Science (ICTS), Bangalore, Dec 2010)}
\end{abstract}

\maketitle

\section{Introduction}
\label{sec:intro}

The Hubbard model is the simplest of a class of models describing electrons moving
on a lattice with repulsive electron-electron interactions. Despite its apparent simplicity, it has become
clear in the past two decades that it can display a very rich phase diagram, with a plethora of interesting phases.
The most common phase is, of course, the Fermi liquid (FL), which is adiabatically connected to the metallic phase
of non-interacting electrons. However, electron-electron interactions can break one or more symmetries of the 
Hamiltonian, and this leads to phases such as antiferromagnets, charge or spin density waves, or superconductors.
Also of great interest are quantum phases which do not break any symmetries, but are nevertheless qualitatively
distinct from the non-interacting electron states: such states are characterized by emergent gauge excitations,
fractionalization of quasiparticle excitations, and non-trivial ground state degeneracies which depend upon the
global topology of the lattice---it is often stated that such states have `topological' order. Finally, there are 
interesting quantum phase transitions between such phases, and such quantum critical points are often described by strongly-coupled
quantum field theories. In some cases, the quantum critical points can broaden into gapless quantum critical phases.

This article will present a pedagogical review of a small sample of this landscape of phases and critical points. 
My aim is to describe the appearance of a variety of non-trivial phases in the simplest possible context. 
For the honeycomb lattice with a density of one electron per site, such phases naturally have low energy excitations
which have a relativistic form at low energies. Consequently, in the vicinity of quantum phase transitions, such phases
and their critical points are amenable to a description by relativistic quantum field theories. In some cases, the critical points
are also conformally invariant, and so are described by conformal field theories (CFTs). In these cases, the AdS/CFT correspondence
can be directly applied, and I will describe the insights that have been gained from such an approach.

However, once we move away from commensurate electron densities, the quantum phases and critical points
of electron lattice models rarely have any relativistic invariance in their low energy theory. I will describe here the simplest examples of 
`topologically ordered' phases at generic electron densities. It is important to note that the lack of relativistic invariance does not rule out 
application of the AdS/CFT correspondence. We can begin from a relativistically invariant gravity dual theory and dope it with charge
carriers by turning on a chemical potential: then even the gravity theory is not relativistically invariant at low energies, and we can
hope to match its low energy physics to a condensed matter system. There has been a large effort to apply the AdS/CFT correspondence along this direction in the past few years. I have discussed some of this work in another
recent review article\cite{statphys}, 
which should be viewed as a companion to the present article. I will also discuss the generic density phases here,
with an emphasis on the general low energy structure of the `fractionalized Fermi liquid' (FL*) phase\cite{ffl1,ffl2}, 
which I believe is closely
related to the generic density phases that have appeared using the AdS/CFT correspondence\cite{ssffl}.
Further details on the connection of these phases of Hubbard-like models and the AdS/CFT correspondence appear in 
a recent paper \cite{liza}.

I begin by introducing the Hubbard model. It is defined by the Hamiltonian
\beq
H = - \sum_{i,j} t_{ij} c^{\dagger}_{i \alpha} c_{j \alpha} + \sum_i \left[- \mu \left( n_{i \uparrow} + n_{i \downarrow} \right) +  U_i \left(n_{i \uparrow} - \frac{1}{2} \right)
\left(n_{i \downarrow} - \frac{1}{2}\right)\right]. \label{h1}
\eeq
Here $c_{i\alpha}$, $\alpha = \uparrow, \downarrow$ are annihilation operators on the site $i$ of a regular lattice,
and $t_{ij}$ is a Hermitian, short-range matrix containing the `hopping matrix elements' which move the electrons
between different lattice sites. The density of electrons is controlled by the chemical potential $\mu$ which couples
to the total electron density, with
\beq
n_{i \uparrow} \equiv c_{i \uparrow}^\dagger c_{i \uparrow} \quad , \quad n_{i \downarrow} \equiv c_{i \downarrow}^\dagger c_{i \downarrow}. \label{h2}
\eeq
The electrons repel each other with an on-site interaction $U_i$; in most cases we will take $U_i = U$ site-independent, but it will also
be useful later to allow for a site-dependent $U_i$. For completeness, we also note the algebra of the fermion  operators:
\bea
c_{i \alpha} c_{j \beta}^\dagger + c_{j \beta}^\dagger c_{i \alpha} &=& \delta_{ij} \delta_{\alpha\beta} \nn
c_{i \alpha} c_{j \beta} + c_{j \beta} c_{i \alpha} &=& 0. \label{h3} 
\eea
The equations (\ref{h1}), (\ref{h2}), and (\ref{h3}) constitute a self-contained and complete mathematical statement of the problem of the landscape of the
Hubbard model. It is remarkable that a problem that is so simple to state has such a rich phase
structure as a function of the lattice choice, the fermion density, and the spatial forms of $t_{ij}$ and $U_i$.

Sections~\ref{sec:afm}, \ref{sec:vbs}, and \ref{sec:su2} will deal exclusively with the honeycomb lattice
at a density of one electron per site (``half-filling''), so that $\left\langle n_{i \uparrow} \right\rangle = 
\left\langle n_{i \downarrow} \right\rangle = 1/2$. The emphasis on the honeycomb lattice is not motivated 
by its particular physical importance (although, it is the lattice of graphene), but by its simplicity as a context for introducing
various technical methods, quantum phases and critical points. In Section~\ref{sec:afm}, we will consider the semi-metal
and the insulating antiferromagnet, and show that a phase transition between them is described by a relativistic field theory, which is a version of the Gross-Neveu-Yukawa model; 
Section~\ref{sec:ads} will use this field theory to present a general discussion of the physics at non-zero
temperatures in the vicinity of a quantum critical point. We will focus on the transport of conserved charges,
and describe insights gained from the AdS/CFT correspondence. 
Section~\ref{sec:vbs} will consider the problem of restoring the spin rotation symmetry
from the antiferromagnet, while remaining in an insulating phase: this will lead to a description in terms of a U(1) gauge theory,
and the appearance of an insulating phase with valence bond solid (VBS) order. Finally, Section~\ref{sec:su2} will combine
all the phases of the half-filled honeycomb lattice discussed so far in a single phase diagram: this will require introduction of
a SU(2) gauge theory. We will find an interesting multi-critical point in Section~\ref{sec:su2}, which has many features in common
with the supersymmetric CFTs studied using the AdS/CFT correspondence.

Sections~\ref{sec:vbs} and~\ref{sec:su2} can be skipped in a first reading, without significant loss of continuity.

We will move away from half-filling in Sections~\ref{sec:tri} and~\ref{sec:ffl}, where we will turn our attention to metallic phases with Fermi surfaces.
Section~\ref{sec:tri} considers the Hubbard model on the triangular lattice, and describes a phase diagram which includes Fermi
liquid (FL) and spin liquid phases. Section~\ref{sec:ffl} extends our discussion to the
the Hubbard model on a bilayer triangular lattice, which has been realized in experiments \cite{saunders} on $^{3}$He.
Here we will present a gauge theory of another metallic phase, the fractionalized Fermi liquid (FL*). We will also discuss the connections
to compressible metallic phases obtained using the AdS/CFT correspondence, complementing the recent discussion in Ref.~\onlinecite{liza}.

\section{Semi-metal and antiferromagnetism on the honeycomb lattice}
\label{sec:afm}

\subsection{Preliminaries}
\label{sec:prelim}

\begin{figure}
\center\includegraphics[width=4in]{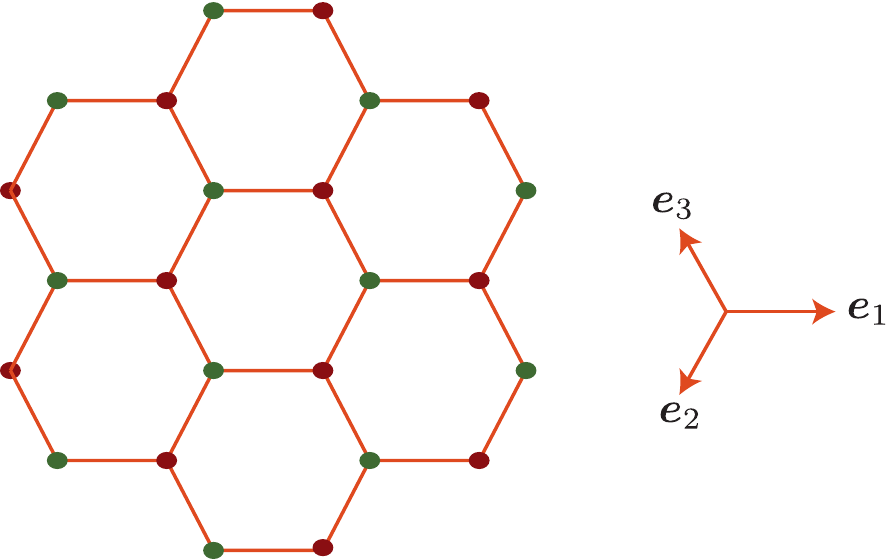}
\caption{The honeycomb lattice with its A (green) and B (red) sublattices}
\label{fig:hclattice}
\end{figure}

We will consider the Hubbard model (\ref{h1}) with the sites $i$ on locations ${\bm r}_i$ on the honeycomb lattice
shown in Fig.~\ref{fig:hclattice}. Here, we set up some notation allowing us to analyze the geometry of this lattice.

We work with a lattice with unit nearest neighbor spacing. We define unit length vectors
which connect nearest-neighbor sites
\begin{equation} 
{\bm e}_1 = (1,0) \quad , \quad {\bm e}_2 = (-1/2, \sqrt{3}/2) \quad , \quad {\bm e}_3 = (-1/2, -\sqrt{3}/2). 
\label{evecs}
\end{equation}
Note that ${\bm e}_i \cdot {\bm e}_j = - 1/2$ for $i \neq j$, and ${\bm e}_1 + {\bm e}_2 + {\bm e}_3 = 0$.
The lattice can be divided into the A and B sublattices, as shown in Fig.~\ref{fig:hclattice}. 
We take the origin of co-ordinates of the lattice at the center of an {\em empty hexagon\/}.
The A sublattice sites closest to the origin are at ${\bm e}_1$, ${\bm e}_2$, and ${\bm e}_3$, while the B sublattice
sites closest to the origin are at  $-{\bm e}_1$, $-{\bm e}_2$, and $-{\bm e}_3$.

The unit cell of the hexagonal lattice contains 2 sites, one each from the A and B
sublattices. These unit cells form a triangular Bravais lattice consisting of the centers
of the hexagons. The triangular lattice points closest to the origin are $\pm ({\bm e}_1 - {\bm e}_2)$,
$\pm ({\bm e}_2 - {\bm e}_3)$, and $\pm ({\bm e}_3 - {\bm e}_1)$. The reciprocal lattice
is a set of wavevectors ${\bm G}$ such that ${\bm G} \cdot {\bm r} = 2 \pi \times$ integer, where ${\bm r}$
is the center of any hexagon of the honeycomb lattice. The reciprocal lattice is also a triangular
lattice, and it consists of the points $\sum_i n_i {\bm G}_i$, where $n_i$ are integers and
\begin{equation}
{\bm G}_1 = \frac{4 \pi}{3} {\bm e}_1 \quad , \quad 
{\bm G}_2 = \frac{4 \pi}{3} {\bm e}_2 \quad , \quad
{\bm G}_3 = \frac{4 \pi}{3} {\bm e}_3 . \label{gvecs}
\end{equation}

\begin{figure}
\center\includegraphics[width=2in]{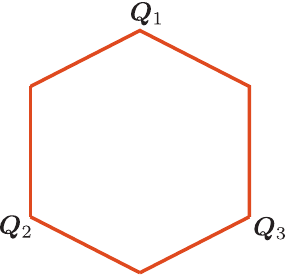}
\caption{The first Brillouin zone of the honeycomb lattice.}
\label{brill}
\end{figure}

The unit cell of the reciprocal lattice is called the first Brillouin zone.
This is a hexagon whose vertices are given by
\begin{equation}
{\bm Q}_1 = \frac{1}{3} ({\bm G}_2 - {\bm G}_3) \quad , \quad 
{\bm Q}_2 = \frac{1}{3} ({\bm G}_3 - {\bm G}_1 )\quad , \quad
{\bm Q}_3 = \frac{1}{3} ({\bm G}_1 - {\bm G}_2 ),  \label{brillQ}
\end{equation}
and $- {\bm Q}_1$, $- {\bm Q}_2$, and $-{\bm Q}_3$; see Fig.~\ref{brill}.
Integrals and sums over momentum space will implicitly extend only over the first Brillouin zone.
This is the `ultraviolet cutoff' imposed by the underlying lattice structure.

We define the Fourier transform of the electrons on the A sublattice by
\begin{equation} c_{A\alpha} ( {\bm k} ) = \frac{1}{\sqrt{\mathcal{N}}} \sum_{i \in A} c_{i \alpha} e^{- i {\bm k} \cdot {\bm r}_i},
\label{fourier}
\end{equation}
where $\mathcal{N}$ is the number of sites on one sublattice;
similarly for $c_{B\alpha}$. Note that $c_{A\alpha} ({\bm k} + {\bm G}) = c_{A \alpha} ({\bm k})$: consequently, sums over momentum
have to be restricted to the first Brillouin zone to avoid double counting. Thus the inverse of Eq.~(\ref{fourier}) sums over
${\bm k}$ in the first Brillouin zone.

\subsection{Semi-metal}
\label{sec:semi}

We begin with free electrons in the honeycomb lattice, $U=0$, with only nearest-neighbor electron hopping
$t_{ij} = t$.
Using Eq.~(\ref{fourier}), we can write the hopping Hamiltonian as
\bea
H_0 &=& -t \sum_{{\bm k}} \left(e^{i {\bm k} \cdot {\bm e}_1} + e^{i {\bm k} \cdot {\bm e}_2}+ e^{i {\bm k} \cdot {\bm e}_3} \right) c_{A \alpha}^\dagger ({\bm k}) c_{B \alpha} ({\bm k}) + \mbox{H.c.} \nn
&~&- \mu \sum_{{\bm k}} \left(  c_{A \alpha}^\dagger ({\bm k}) c_{A \alpha} ({\bm k}) + c_{B \alpha}^\dagger ({\bm k}) c_{B \alpha} ({\bm k}) 
\right)
\eea
We introduce Pauli matrices $\tau^a$ ($a = x,y,z$) which act on the $A$, $B$ sublattice space; then this Hamiltonian can be written as
\begin{eqnarray}
H_0 &=& \sum_{{\bm k}} c^{\dagger} ({\bm k}) \Bigl[ - \mu  -t \Bigl(\cos({\bm k} \cdot {\bm e}_1) + \cos({\bm k} \cdot {\bm e}_2)
+ \cos({\bm k} \cdot {\bm e}_3) \Bigr) \tau^x \nonumber \\
&~&~~~~~~~+ t \Bigl(\sin({\bm k} \cdot {\bm e}_1) + \sin({\bm k} \cdot {\bm e}_2)
+ \sin ({\bm k} \cdot {\bm e}_3) \Bigr) \tau^y \Bigr] c ({\bm k} ), \label{hlat}
\end{eqnarray}
where the sublattice and spin indices on the electrons are now implicit: the $c({\bm k})$ are 4-component fermion operators.

\begin{figure}
\center\includegraphics[width=3in]{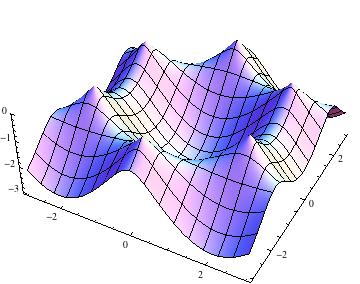}
\caption{The lower band of the dispersion in Eq.~(\ref{spec}) for $\mu=0$}
\label{disp}
\end{figure}

The energy eigenvalues are easily determined to be
\beq -\mu \pm \left| e^{i {\bm k} \cdot {\bm e}_1} + e^{i {\bm k} \cdot {\bm e}_2}+ e^{i {\bm k} \cdot {\bm e}_3} \right| \label{spec}
\eeq
and these are plotted in Fig.~\ref{disp}. 
At half-filling, exactly half the states should be occupied in the ground state,
and for the spectrum in Eq.~(\ref{spec}) this is achieved at $\mu=0$.

A crucial feature of any metallic state is the Fermi surface: this is boundary between the occupied and empty
states in momentum space. In two spatial dimensions, this boundary is generically a line in momentum space,
and this is the case for the dispersion in Eq.~(\ref{spec}) for $\mu \neq 0$. However, for the $\mu=0$, the 
honeycomb lattice has the special property that the occupied and empty states meet only at a discrete set of single points
in momentum space: this should be clear from the dispersion plotted in Fig.~\ref{disp}. Only 2 of these points are distinct, in that
they are not separated by a reciprocal lattice vector ${\bm G}$. So the 
 half-filled
honeycomb lattice has 2 `Fermi points', and realizes a `semi-metal' phase. The low energy excitations of the semi-metal consist
of particles and holes across the Fermi point, and these have a lower density of states than in a metallic phase with a Fermi line.
We also note that the Fermi-point structure is protected by a sublattice exchange symmetry: it is not special to the nearest-neighbor
hopping model, and it also survives the inclusion of electron-electron interactions.

We obtain a very useful, and universal, theory for the low energy excitations of the semi-metal by expanding (\ref{hlat}) in the vicinity
of the Fermi points. The distinct Fermi points are present at ${\bm Q}_1$ and $-{\bm Q}_1$; all other Fermi points are separated from
these two points by a reciprocal lattice vector ${\bm G}$. So we define continuum Fermi field which reside in `valleys' in the vicinity
of these points by 
\begin{eqnarray}
C_{A1\alpha} ({\bm k} ) &=& \sqrt{A} \,
c_{A\alpha} ({\bm Q}_1 + {\bm k} ) \nonumber \\
C_{A2\alpha} ({\bm k} ) &=& \sqrt{A}\,
c_{A\alpha} (-{\bm Q}_1 + {\bm k} ) \nonumber \\
C_{B1\alpha} ({\bm k} ) &=& \sqrt{A}\,
c_{B\alpha} ({\bm Q}_1 + {\bm k} ) \nonumber \\
C_{B2\alpha} ({\bm k} ) &=& \sqrt{A}\,
c_{B\alpha} (-{\bm Q}_1 + {\bm k} ), \label{cont}
\end{eqnarray}
where $A$ is the total area of the honeycomb lattice, and the momentum ${\bm k}$ is small.
The field $C$ is a 8-component continuum canonical Fermi field: the components correspond to 
spin ($\uparrow$, $\downarrow$), sublattice ($A$, $B$), and valley ($1,2$) indices. We will also use Pauli
matrices which act on the spin ($\sigma^a$), sublattice ($\tau^a$), and valley ($\rho^a$) space.

Inserting Eq.~(\ref{cont}) into Eq.~(\ref{hlat}), we obtain the continuum Hamiltonian
\begin{equation}
H_0 =  \int \frac{d^2 k}{4 \pi^2} C^{\dagger} ({\bm k} ) \Bigl( v \tau^y k_x + v \tau^x \rho^z k_y \Bigr) C ({\bm k}), \label{ham}
\end{equation}
where $v=3t/2$. From now on we rescale time to set $v=1$. Diagonalizing Eq.~(\ref{ham}), we obtain the relativistic spectrum
\beq \pm \sqrt{k_x^2 + k_y^2}, \eeq which corresponds to the values of Eq.~(\ref{spec}) near the Fermi points.

The relativistic structure of $H_0$ can be made explicit by rewriting it as the Lagrangian of massless Dirac fermions. 
Define $\overline{C} = C^{\dagger} \rho^z \tau^z$. Then we can write the Euclidean time ($\tau$)
Lagrangian density of the semi-metal phase as 
\begin{equation}
\mathcal{L}_0 =  \overline{C} \left( \partial_\tau \gamma_0 + \partial_x \gamma_1 + \partial_y \gamma_2 \right) C
\end{equation} 
where $\omega$ is the frequency associated with imaginary time, and the Dirac $\gamma$ matrices are
\begin{equation}
\gamma_0 = - \rho^z \tau^z \quad \gamma_1 = \rho^z \tau^x \quad \gamma_2 = - \tau^y . \label{defgamma}
\end{equation}
In addition to relativistic invariance, this form makes it clear the free-fermion Lagrangian has a large group
of `flavor' symmetries that acts on the 8$\times$8 fermion space and commute with the $\gamma$ matrices. Most of these 
symmetries are not obeyed by higher-order gradients, or by fermion interaction terms which descend from the Hubbard model.

Let us now turn on a small repulsion, $U$, between the fermions in the semi-metal. Because of the point-like
nature of the Fermi surface, it is easier to determine the consequences of this interaction here than in a metallic phase
with a Fermi line of gapless excitations. We can use traditional renormalization group (RG) methods to conclude
that a weak $U$ is irrelevant in the infrared: the computation is left as an exercise below. Consequently, the semi-metal
state is a stable phase which is present over a finite range of parameters.\\~\\
\noindent
{\bf Exercise:} Observe that $\mathcal{L}_0$ is invariant under the scaling transformation
$x' = x e^{-\ell}$ and $\tau' = \tau e^{- \ell}$. Write the Hubbard interaction $U$ in terms of the Dirac fermions,
and show that it has the tree-level scaling transformation $U' = U e^{-\ell}$. So argue that all short-range interactions
are {\em irrelevant\/} in the Dirac semi-metal phase.\\

\subsection{Antiferromagnet}
\label{sec:afm2}

Although a small $U$ is irrelevant, new phases can and do appear at large $U$. 
To see this, let us return to the lattice Hubbard model in Eq.~(\ref{h1}), and consider
the limit of large $U_i = U $. We will assume $\mu=0$ and half-filling in the remainder of this section.

At $U=\infty$, the eigenstates are simple products over the states on each site. Each site has 4 states:
\beq
|0 \rangle \quad,\quad c_{i \uparrow}^\dagger |0 \rangle \quad , \quad c_{i \downarrow}^\dagger |0 \rangle \quad , \quad
c_{i \uparrow}^\dagger c_{i \downarrow}^\dagger |0 \rangle, \label{4state}
\eeq
where $|0 \rangle$ is the empty state. The energies of these states are $U/4$, $-U/4$, $-U/4$, and $U/4$ respectively.
Thus the ground state on each site is doubly-degenerate, corresponding to the spin-up and spin-down states of a single
electron. The lattice model has a degeneracy of $2^{2 \mathcal{N}}$, and so a non-zero entropy density
(recall that $\mathcal{N}$ is the number of sites on one sublattice).

Any small perturbation away from the $U=\infty$ limit is likely to lift this exponential large degeneracy.
So we need to account for the electron hopping $t$. At first order, electron hopping moves an electron from 
one singly-occupied site to another, yielding a final state with one empty and one doubly occupied site.
This final state has an energy $U$ higher than the initial state, and so is not part of the low energy manifold.
So by the rules of degenerate perturbation theory, there is no correction to the energy of all the $2^{2 \mathcal{N}}$
ground states at first order in $t$.

At second order in $t$, we have to use the effective Hamiltonian method. This performs a canonical transformation to
eliminate the couplings from the ground states to all the states excited by energy $U$, while obtaining a modified
Hamiltonian which acts on the $2^{2 \mathcal{N}}$ ground states. This method is described in 
text books on quantum mechanics, and we leave its application here as an exercise. The resulting effective
Hamiltonian is the Heisenberg antiferromagnet:
\begin{equation}
H_J =  \sum_{i<j} J_{ij} S^a_i S^a_j \quad, \quad J_{ij} = \frac{4 t_{ij}^2}{U}, \label{HJ}
\end{equation}
where $J_{ij}$ is the exchange interaction and $S^a_i$ are the spin operators on site $i$
\beq S^a_i = \frac{1}{2} c_{i\alpha}^\dagger \sigma^a_{\alpha\beta} c_{i \beta}. \eeq
Note that these spin operators preserve the electron occupation number on every site,
and so act within the subspace of the $2^{2 \mathcal{N}}$ low energy states.
The Hamiltonian $H_J$ lifts the macroscopic degeneracy, and the entropy density of the new
ground state will be zero.\\~\\
{\bf Exercise:} Use the effective Hamiltonian method described in Ref.~\onlinecite{cohen} to obtain Eq.~(\ref{HJ}).
At second order in $U$, it is sufficient to consider the 2-site Hubbard model. This has a total of 16 states,
and 4 ground states at $U=\infty$. Derive the effective Hamiltonian which acts on these 4 states at
order $t^2$.\\

\begin{figure}
\center\includegraphics[width=2.2in]{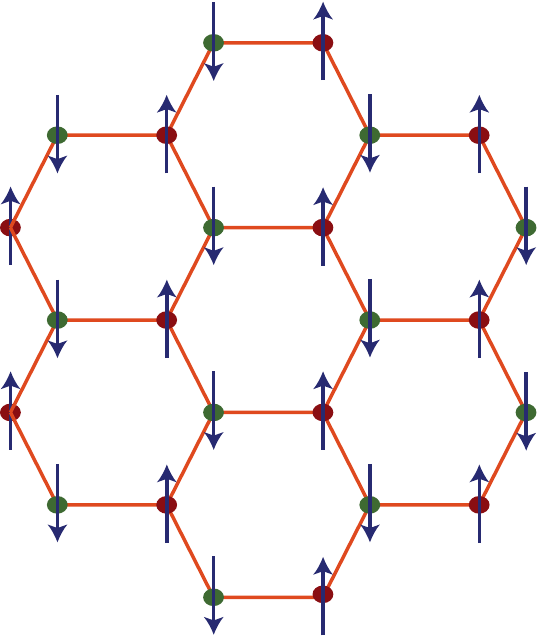}
\caption{The large $U$ state with antiferromagnetic (N\'eel) order.}
\label{fig:neel}
\end{figure}

Although we cannot compute the exact ground state of $H_J$ on the honeycomb lattice with nearest-neighbor
exchange, numerical studies\cite{meng} leave little doubt
to its basic structure. The ground state is adiabatically connected to that obtained by treating the $S^a_i$ 
as classical vectors in spin space: it has antiferromagnetic (or N\'eel) order which breaks the global SU(2)
spin rotation symmetry, by a spontaneous polarization of the spins on opposite orientations on the two
sublattices
\beq \eta_i \left \langle S^a_i \right\rangle = N^a, \eeq
where $\eta_i = 1$ ($\eta_i = -1$) on sublattice A ($B$), and $N^a$ is the vector N\'eel order parameter;
see Fig.~\ref{fig:neel}.
Classically this state minimizes the exchange coupling in Eq.~(\ref{HJ}) because $J_{ij} > 0$. Quantum fluctuations for 
spin $S=1/2$ reduce the spontaneous moment from its classical value, but a non-zero moment remains on the
honeycomb lattice.

What is the electronic excitation spectrum in the antiferromagnet? To determine this, it is useful to write
the N\'eel order parameter in terms of the continuum Dirac fields introduced in Section~\ref{sec:semi}.
We observe
\beq \sum_i \eta_i S_i^a = \sum_{{\bm k}} \left( c_{A \alpha}^\dagger \sigma^a_{\alpha\beta} c_{A \beta} - c_{B \alpha}^\dagger \sigma^a_{\alpha\beta} c_{B \beta} \right) = \int \frac{d^2 k}{4 \pi^2} C^\dagger \tau^z \sigma^a C \label{cneel}
\eeq
Thus the N\'eel order parameter $N^a$ is given by the fermion bilinear 
\beq
N^a = \left\langle C^\dagger \tau^z \sigma^a C \right\rangle =  \left\langle \overline{C} \rho^z \sigma^a C \right\rangle,\eeq
and the vacuum expectation value (VEV) is non-zero in the antiferromagnet. 
We can expect that electron-electron interactions will induce a coupling between the fermion excitations and this VEV
in the low energy Hamiltonian for the N\'eel phase. Choosing N\'eel ordering in the $z$ direction
\beq N^a = N_0 \delta_{az}, \eeq  we anticipate that $H_0$ in Eq.~(\ref{ham}) is modified
in the N\'eel phase to
\beq
H_N = \int \frac{d^2 k}{4 \pi^2} C^{\dagger} ({\bm k} ) \Bigl(  \tau^y k_x +  \tau^x \rho^z k_y - \lambda N_0 \tau^z \sigma^z \Bigr) C ({\bm k}), \label{hamn}
\end{equation}
where $\lambda$ is a coupling determined by the electron interactions, and we have assumed N\'eel order polarized in the $z$
direction. This effective Hamiltonian will be explicitly derived in the next subsection.
We can now easily diagonalize $H_N$ to deduce that the electronic excitations have energy
\beq \pm \sqrt{ k_x^2 + k_y^2 + \lambda^2 N_0^2}. \label{massivedirac} \eeq
This is the spectrum of massive Dirac fermions. So the Fermi point has disappeared, and an energy gap has
opened in the fermion excitation spectrum. In condensed matter language, the phase with antiferromagnetic order
is an insulator, and not a semi-metal: transmission of electronic charge will require creation of gapped particle
and hole excitations. 

\subsection{Quantum phase transition}
\label{sec:qpt}

We have now described a semi-metal phase for small $U$, and an antiferromagnetic insulator for large $U$.
Both are robust phases, whose existence has been reliably established. We now consider connecting
these two phases at intermediate values of $U$. This is a complex subject, and careful numerical studies
are only just emerging for the model with nearest-neighbor hopping\cite{meng}. It is already clear, however, that by
varying the form of the microscopic coupling we can obtain a rich variety of intermediate phases\cite{hermelehc,fwang,ran1,ran2,cenke1,cenke2}.
In the present subsection we consider the simplest possibility: there are no new intermediate phases,
and only a direct quantum phase transition between the semi-metal and the 
antiferromagnetic insulator\cite{herbut0,herbut1,herbut2}.

We can derive the field theory for this direct transition either by symmetry considerations, or by an 
explicit derivation from the Hubbard model. Let us initially follow the second route.
We start with the Hubbard Hamiltonian in Eq.~(\ref{h1}), 
 use the operator identity (valid on each site $i$):
\begin{equation}
U \left(n_\uparrow - \frac{1}{2} \right) \left(n_\downarrow - \frac{1}{2} \right) = -\frac{2U}{3} S^{a2} + \frac{U}{4}.
\end{equation}
Then, in the fermion coherent state path integral for the Hubbard model, we apply a `Hubbard-Stratonovich' transformation
to the interaction term; this amounts to using the identity
\begin{eqnarray} 
&& \exp \left( \frac{2U}{3} \sum_i \int d \tau S^{a2}_i \right) \nn
&&~~~= \int \mathcal{D} X^a_i (\tau) \exp
\left( - \sum_i \int d \tau \left[ \frac{3}{8} X^{a2}_i - \sqrt{U} X^a_i S^a_i \right] \right) \label{hs}
\end{eqnarray}
The fermion path integral is now a bilinear in the fermions, and we can, at least formally, integrate out
the fermions in the form of a functional determinant. We imagine doing this temporarily, and then 
look for the saddle point of the resulting effective
action for the $X^a_i$. At the saddle-point we find that the lowest energy is achieved
when the vector has opposite orientations on the A and B sublattices. Anticipating this, we
look for a continuum limit in terms of a field $\varphi^a$ where
\begin{equation}
X^a_i =  \eta_i \varphi^a 
\end{equation}
Using Eq.~(\ref{cneel}), the
continuum limit of the coupling between the field $\varphi^a$ and the fermions in Eq.~(\ref{hs}) 
is given by
\beq X^a_i c_{i \alpha}^\dagger \sigma^a_{\alpha\beta} c_{i \beta} =
 \varphi^a C^\dagger \tau^z \sigma^a C =  \varphi^a \overline{C} \rho^z \sigma^a C \label{phiC}
 \eeq
From this it is clear that $\varphi^a$ is a dynamical quantum field which represents the fluctuations of the local
N\'eel order, and \beq \langle \varphi^a \rangle \propto N^a. \eeq 
 
Now we can take the continuum limit of all the terms in the coherent state path integral for the lattice Hubbard
model and obtain the following continuum Lagrangian density  
\begin{equation}
\mathcal{L} =   \overline{C} \gamma_\mu \partial_\mu C + \frac{1}{2} \left[ \left( \partial_\mu \varphi^a \right)^2 
+ s \varphi^{a2} \right]  + \frac{u}{24}  \left( \varphi^{a2} \right)^2 - \lambda \varphi^a \overline{C} \rho^z \sigma^a C \label{GN}
\end{equation}
This is a relativistic quantum field theory for the 8-component fermion field $C$ and the 3-component real scalar $\varphi^a$,
related to the Gross-Neveu-Yukawa model.
We have included gradient terms and quartic in the Lagrangian for $\varphi^a$: these are not present in the derivation
outlined above from the lattice Hubbard model, but are clearly induced by higher energy fermions are integrated out.
The Lagrangian includes various phenomenological couplings constants ($s$, $u$, $\lambda$); as these constants are
varied, $\mathcal{L}$ can describe {\em both\/} the semi-metal and insulating antiferromagnet phases,
and also the quantum critical point between them.

Note that the matrix $\rho^z \sigma^a$ commutes with all the $\gamma_\mu$;
hence $\rho^z \sigma^a$ is a matrix in ``flavor'' space. So if we consider $C$ as 2-component Dirac
fermions, then these Dirac fermions carry an additional 4-component flavor index.

The semi-metal phase is the one where $\varphi^a$ has vanishing VEV. In mean-field theory, this appears
for $s>0$. The $\varphi^a$ excitations are then massive, and these constitute a triplet of gapped `spin-excitons' 
associated with fluctuations of the local antiferromagnetic order. The Dirac fermions are massless,
and represent the Fermi point excitations of the semi-metal.

The N\'eel phase has a non-zero VEV, $\langle \varphi^a \rangle \neq 0$, and appears in mean-field
theory for $s<0$. Here the Dirac fermions acquire a gap, indicating that the Fermi point has vanished,
and we are now in an insulating phase. The fluctuations of $\varphi$ are a doublet of Goldstone
modes (`spin waves') and a longitudinal massive Higgs boson.

Finally, we are ready to address the quantum critical point between these phases.
In mean-field theory, this transition occurs at $s=0$. As is customary in condensed matter physics,
it is useful to carry out an RG analysis near this point. The tree-level analysis is carried out in the following
exercise.\\~\\
\noindent
{\bf Exercise:} Perform a tree-level RG transformation on $\mathcal{L}$. The quadratic gradient terms are invariant
under $C' = C e^{\ell}$ and $\varphi' = \varphi e^{\ell/2}$. Show that this leads to $s' = s e^{2 \ell}$. Thus $s$ is 
a relevant perturbation which drives the system into either the semi-metal or antiferromagnetic insulator.
The quantum critical point is reached by tuning $s$ to its critical value ($=0$ at tree level). Show that the couplings
$u$ and $\lambda$ are both relevant perturbations at this critical point. Thus, while interactions are irrelevant
in the Dirac semi-metal (and in the insulator), they are strongly relevant at the quantum-critical point.\\

Further study of this quantum critical point requires a RG analysis which goes beyond tree-level. Such an 
analysis can be controlled in an expansion in $1/N$ (where $N$ is the number of fermion flavors)
or $(3-d)$ (where $d$ is the spatial dimensionality. For reviews see Ref.~\onlinecite{altenberg}
or Chapter 17 of Ref.~\onlinecite{ssbook2}.  
The main conclusion of such analyses is that there is an RG fixed point at which the $\varphi^{a2}$ is the only 
relevant perturbation. Non-linearities such as $\lambda$ and $u$ all reach stable fixed point values of order
unity. This non-trivial fixed point implies that the physics of the quantum critical point is highly non-trivial and strongly
coupled. The RG fixed point is scale- and relativistic-invariant, and this implies that it is also conformally invariant.
Thus the quantum critical point is described by a CFT in 2+1 spacetime dimensions: a CFT3.

We will not describe the critical theory in any detail here.
However, we will note some important characteristics of correlation functions at the quantum critical point.
The electron Green's function has the following structure
\begin{equation}
\left\langle C (k, \omega) ; C^\dagger (k, \omega) \right\rangle \sim 
\frac{ i \omega +  k_x \tau^y +  k_y \tau^x \rho^z }{(\omega^2 +  k_x^2 +  k_y^2)^{1- \eta/2}}
\end{equation}
where $\eta>0$ is the {\em anomalous dimension} of the fermion. This leads to a fermion 
spectral density which has no quasiparticle pole: thus the quantum critical point has no well-defined quasiparticle excitations.
This distinguishes it from both the semi-metal and insulating antiferromagnetic phases that flank it on either side: both had
excitations with infinitely-sharp quasiparticle peaks. Similar anomalous dimensions appear in the correlations
of the bosonic order parameter $\varphi^a$.

\subsection{Quantum impurities}
\label{sec:qimp}

We briefly note the physics of quantum impurities in the honeycomb lattice, which were discussed more completely
in the companion review\cite{statphys}. The translational invariance of the honeycomb lattice
will be broken {\em only\/} in this subsection.

\begin{figure}
\center\includegraphics[width=2.2in]{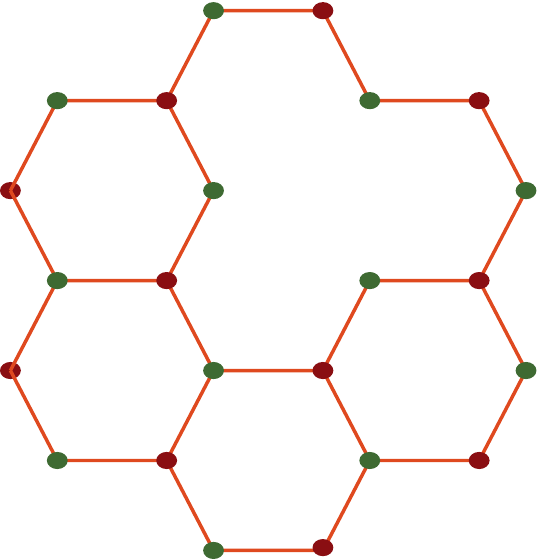}
\caption{The honeycomb lattice with a vacancy.}
\label{fig:vac}
\end{figure}

Imagine removing a single atom from the honeycomb lattice, as shown in Fig.~\ref{fig:vac}
At $U=0$, the electronic spectrum of such a lattice was described in Ref.~\onlinecite{castro}.
They showed that there was a quasi-localized state in the vicinity of the impurity exactly at zero
energy, which decayed as $1/r$ at a distance $r$ from the impurity. Upon including $U$, we expect this
state to capture a single electron whose $S=1/2$ spin interacts with the bulk excitations. 

Let us represent the impurity by a localized fermion operator $\chi_\alpha (\tau)$. Note that this fermion
has no dependence upon the spatial co-ordinate ${\bm r}$, and is a function only of the time $\tau$.
Now we can couple this fermion to the bulk excitations which were described by $\mathcal{L}$ in
Eq.~(\ref{GN}) in the vicinity of the semi-metal to antiferromagnetic insulator transition. The full system is
described by the Lagrangian $\mathcal{L}+\mathcal{L}_{\rm imp}$ where
\beq
\mathcal{L}_{\rm imp} = \chi_\alpha^\dagger \frac{\partial \chi_\alpha}{\partial \tau} - h \, \chi_\alpha^\dagger \sigma^a_{\alpha\beta} \chi_\beta \, \varphi^a ({\bm r=0}, \tau);
\label{limp}
\eeq
note that whereas the Lagrangian $\mathcal{L}$ is integrated over spacetime, the Lagrangian $\mathcal{L}_{\rm imp}$
is only integrated over time. There are many possible couplings between the impurity 
and bulk degrees of freedom which are allowed by the symmetry of the problem, but we have only included
a single one. This is easily seen to be the only term which is relevant under the RG which applies
in the vicinity of the bulk quantum phase transition. 

The RG flow of the bulk-impurity coupling $h$ was described in Refs.~\onlinecite{eps1,eps2,eps3}. 
It was found that $h$ approached
a fixed-point coupling in the infrared, just like the couplings $u$ and $\lambda$ in $\mathcal{L}$.
Thus no new couplings are necessary to describe the low energy physics of the impurity provided
we are not too far from the semi-metal to antiferromagnetic insulator quantum critical point.

Further details of the impurity dynamics may be found in 
the companion review\cite{statphys}, where it was described for a closely
related bulk quantum critical point. A close analogy was also drawn between these impurity
problems and defects in super Yang-Mills theories;
the latter can be solved\cite{kachru1} by intersecting brane models in string theory, and led to a
description of the impurity criticality using a AdS$_2$ geometry.

\subsection{Electrical transport}
\label{sec:elec}

We now revert to the honeycomb lattice without impurities.

An important set of observables which do {\em not\/} acquire anomalous dimensions at the quantum critical
point are the currents associated with global conservation laws. As the simplest example here, let us consider
correlations of the conserved electric charge of the electrons, and the associated electrical conductivity $\sigma$.
At zero temperature ($T=0$), we have $\sigma = 0$ in the insulator, while the semi-metal and the quantum critical 
point have finite non-zero
values of $\sigma$, as we will now see. 

The conserved electrical current is
\beq J_\mu = -i \overline{C} \gamma_\mu C. \eeq
Let us compute its two-point correlator, $K_{\mu\nu} (k)$ at a spacetime momentum $k_\mu$. 
At leading order, this is given by a one fermion loop diagram which evaluates to
\bea K_{\mu\nu} (k) &=& \int \frac{d^3 p}{8 \pi^3} \frac{ \mbox{Tr} \left[ \gamma_\mu ( i \gamma_\lambda p_\lambda 
+ m \rho^z \sigma^z )
\gamma_\nu ( i \gamma_\delta ( k_\delta + p_\delta) + m \rho^z \sigma^z )\right] }{(p^2 + m^2) ((p+k)^2 + m^2)} \nn
&=&- \frac{2}{\pi} \left( \delta_{\mu\nu} - \frac{k_\mu k_\nu}{k^2} \right) 
\int_0^1 dx \frac{k^2 x (1-x)}{\sqrt{m^2 + k^2 x (1-x)}},
\label{Kmn}
\eea
where the mass $m=0$ in the semi-metal and at the quantum critical point, while $m = |\lambda N_0|$ in
the insulator. Note that the current correlation is purely transverse, and this follows from the requirement
of current conservation
\beq k_\mu K_{\mu\nu} = 0. \label{ccons} \eeq
Of particular interest to us is the $K_{00}$ component, after analytic continuation to
Minkowski space where the spacetime momentum $k_\mu$ is replaced by $(\omega, k)$.
The conductivity is obtained from this correlator via the Kubo formula
\beq
\sigma (\omega) = \lim_{k \rightarrow 0} \frac{-i \omega}{k^2} K_{00} (\omega, k). \label{kubo}
\eeq

In the insulator, where $m > 0$, analysis of the integrand in Eq.~(\ref{Kmn}) shows that that the spectral weight of the 
density correlator has a gap of $2m$ at $k=0$, and the conductivity in Eq.~(\ref{kubo}) vanishes.
These properties are as expected in any insulator.

In the metal, and at the critical point, where $m=0$, the fermionic spectrum is gapless, and so is
that of the charge correlator. The density correlator in Eq.~(\ref{Kmn}) and
the conductivity in Eq.~(\ref{kubo}) evaluate to the simple universal results
\bea
K_{00} (\omega, k) &=& \frac{1}{4} \frac{k^2}{\sqrt{k^2 - \omega^2}} \nn
\sigma (\omega) &=& \frac{1}{4}. \label{univ}
\eea

How about beyond the one-loop results? The insulator maintains a gap to charged excitations, 
and so the conductivity remains at zero. In the semi-metal, the fermions are gapless, but they couple only
to the gapped fluctuations of the N\'eel order $\varphi^a$. Examination of the perturbation theory
shows that these have no effect on the current correlators at small momenta and frequency, 
and so the results in Eq.~(\ref{univ}) are {\em exact\/} in the limit of small $\omega$ and $k$ in the 
semi-metal.

At the quantum critical point, we have to consider the strong critical fluctuations associated
with fixed-point values of the Yukawa coupling $\lambda$ and the quartic bosonic interaction $u$.
These can be examined in the $(3-d)$ or the $1/N$ expansion, and require evaulation of multi-loop
diagrams. We will not describe the computations here, but note a remarkable feature: all divergences
associated with the critical fluctuations cancel, and the final result is universal. The values of none of the couplings
of the Lagrangian in Eq.~(\ref{GN}) matters because these are all pinned by the RG fixed point.
There are no anomalous dimensions, and the results in Eq.~(\ref{univ}) generalize to 
\bea
K_{00} (\omega, k) &=& \mathcal{K} \frac{k^2}{\sqrt{k^2 - \omega^2}} \nn
\sigma (\omega) &=& \mathcal{K}, \label{univ2}
\eea
where $\mathcal{K}$ is a universal number dependent only upon the universality class of the 
quantum critical point. The value of the $\mathcal{K}$ for the Gross-Neveu-Yukawa model in Eq.~(\ref{GN}) is not
known exactly, but can be estimated by computations in the $(3-d)$ or $1/N$ expansions.

\section{Non-zero temperatures and the AdS/CFT correspondence}
\label{sec:ads}

We begin by some general remarks on the influence of a non-zero temperature, $T$, on a
generic, strongly-coupled quantum critical point. Let us consider a quantum-critical point which has
only a single relevant perturbation, $s$, as is the case for the model in Eq.~(\ref{GN}) (the generalization
to several relevant perturbations is immediate). So near the quantum critical point, the RG flow is described by
\beq \frac{ds}{d\ell} = \frac{1}{\nu} s. \label{sflow} \eeq
In standard condensed matter notation, the eigenvalue of the relevant flow is written in terms of $\nu$, the correlation
length exponent. Now in the quantum field theory in Euclidean time, a non-zero $T$ corresponds to placing
the theory on a cylinder of circumference $1/T$. Such a finite size is clearly relevant in the infrared,
and also indicates that $1/T$ transforms just like the temporal length under the RG. We write this as
\beq \frac{d T}{d \ell} = z T, \label{Tflow} \eeq
where $z$ is the dynamic critical exponent. All the theories for the honeycomb lattice at half filling
have $z=1$, but we allow $z$ to be arbitrary here.

\begin{figure}
\center\includegraphics[width=4in]{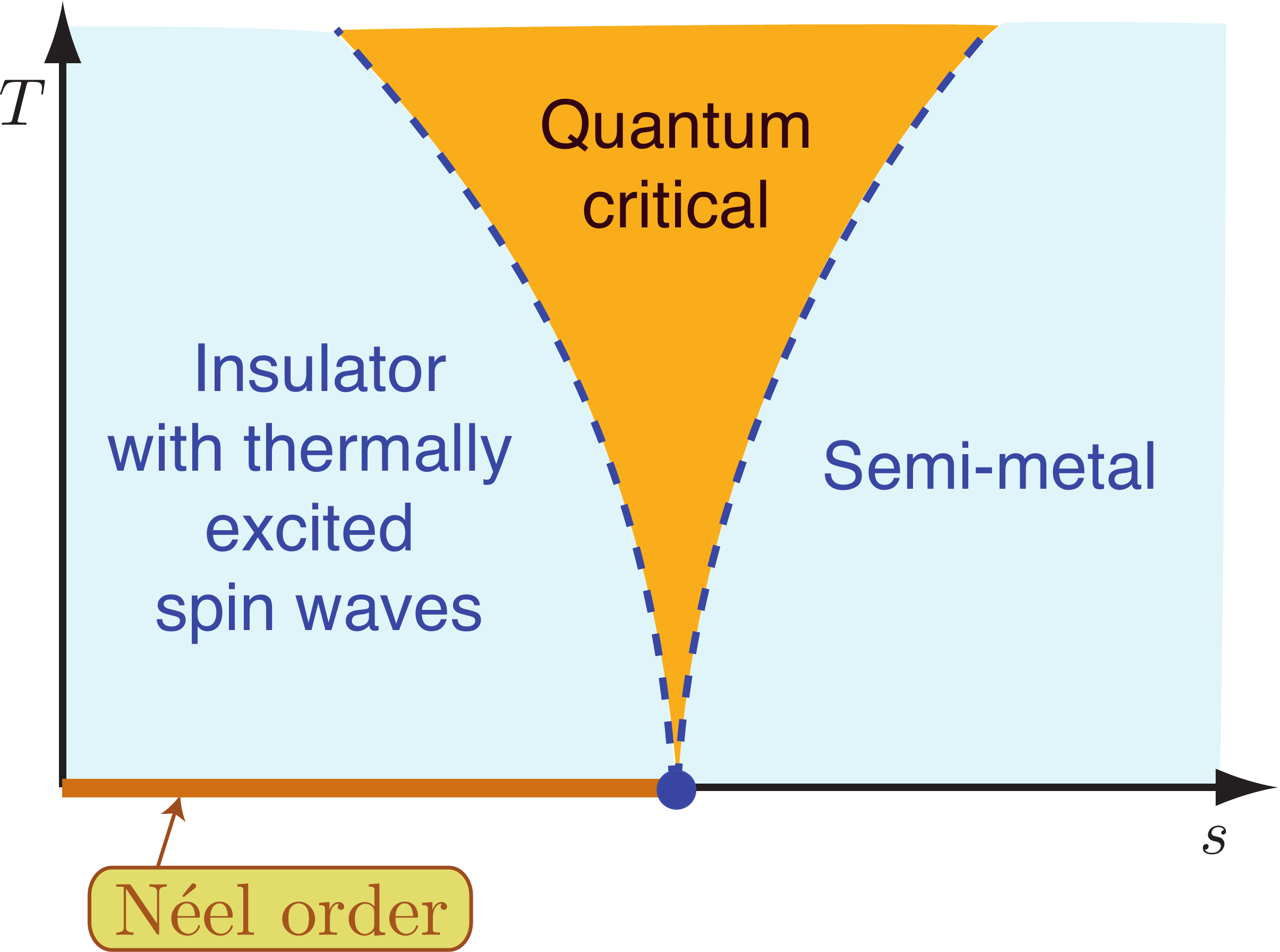}
\caption{Canonical quantum critical crossover phase diagram. The dashed lines occur
for $T \sim |s|^{z \nu}$, and indicate
crossovers between the orange- and blue-shaded regions which are generic
from any strongly-coupled quantum critical point. Specific features of the blue-shaded regions
for the theory Eq.~(\ref{GN}) of the transition from the semi-metal to the 
N\'eel-ordered insulator are also indicated. The N\'eel order vanishes for any $T>0$
because non-Abelian continuous symmetries cannot be broken in two spatial dimensions.}
\label{qccross}
\end{figure}

Eqs.~(\ref{sflow}) and (\ref{Tflow}) are of course trivial to integrate
\beq s(\ell) = s e^{\ell/\nu} \quad, \quad T(\ell) = T e^{z \ell}, \eeq
but the results teach us an important lesson which is summarized in the 
canonical quantum-critical phase diagram shown in Fig.~\ref{qccross}.
We ask the question: which of the relevant perturbations, $s$ or $T$, is more important?
To answer this question, we integrate the RG equations to a scale  $\ell = \ell^\ast$ until
the winner reaches a value of unity. This allows us to conclude that for $T > |s|^{z \nu}$,
thermal effects are more important than any deviation of the coupling from
the RG fixed point. Conversely, for $T< |s|^{z \nu}$ the couplings flow far from
the critical fixed point before any thermally excited states need be considered.
These considerations lead to the two distinct regimes show in Fig.~\ref{qccross}.

In the blue-colored regimes of Fig.~\ref{qccross}, where $T < |s|^{z\nu}$, the physics
of the two non-critical phases dominates. For the model of Eq.~(\ref{GN}), these
are the semi-metal or antiferromagnetic insulator phases, both of which have
well-defined quasiparticle excitations. Consequently, the long-time dynamics can be
written using quasi-classical models of the interactions of these quasiparticles.

Our interest here is primarily in the orange-colored regime of quantum criticality,
$T > |s|^{z \nu}$. Here $T$ is the primary perturbation to the quantum critical theory.
The deviation of the couplings from $T=0$ RG fixed point is unimportant, and the
system behaves as it is described by the universal quantum-critical Lagrangian
in the entire regime. For the relativistic model considered here, the strongly coupled
CFT describes the dynamics of the orange-colored region.

It has been argued \cite{ssbook2} that a central general property of quantum critical dynamics is the short time
over which the system relaxes back to thermal equilibrium. We imagine perturbing
the system away from equilibrium, and measuring the time, $\tau_{\rm eq}$ over
which it relaxes back to local equilibrium (the adjective `local' implies that we exclude
diffusion of globally conserved charges which can take a long time to reach equilibrium
across the entire system). In the regime of strongly-coupled quantum criticality this is given by
\beq \tau_{\rm eq} = \mathcal{C} \frac{\hbar}{k_B T} \label{relax} \eeq
where $\mathcal{C}$ is a universal number dependent only upon the universality class
of the transition, and precise definition used for $\tau_{\rm eq}$. Furthermore, in all
other regimes, the value of $\tau_{\rm eq}$ is parametrically larger than the value
in Eq.~(\ref{relax}). Thus quantum criticality is described by a quantum fluid with the {\em shortest possible\/}
thermal equilibration time. This characteristic makes it a ``nearly perfect'' fluid.

It is important to note that our discussion above does {\em not\/} apply to CFTs in
1+1 dimensions. These are integrable systems, whose long-time dynamics is non-generic
and does not generalize to higher dimensions.

\subsection{Quantum critical transport}
\label{sec:qct}

Let us explore the ideas above by examining the behavior of the electron conductivity
of the model in Eq.~(\ref{GN}) in the quantum-critical regime.
At one-loop order, we can set $m=0$, and then repeat the computation in Eq.~(\ref{Kmn})
at $T>0$. This only requires replacing the integral over the loop frequency by a summation
over the Matsubara frequencies, which are quantized by odd multiples of $\pi T$.
Such a computation, via Eq.~(\ref{kubo}) leads to the conductivity \cite{ssqhe}
\bea
\mbox{Re}[\sigma (\omega)]  &=& (2 T \ln 2) \, \delta (\omega) + \frac{1}{4} \tanh \left( \frac{|\omega|}{4T} \right) \nonumber \\
\mbox{Im}[\sigma(\omega)] &=& \int_{-\infty}^{\infty} \frac{d \Omega}{\pi} \, \mathcal{P} \, 
\left( \frac{\mbox{Re}[\sigma (\Omega)] - 1/4}{\omega-\Omega} \right) ,
\label{drude}
\eea
where $\mathcal{P}$ is the principal part.
Note that this reduces to Eq.~(\ref{univ}) in the limit $\omega \gg T$.
However, the most important new feature of Eq.~(\ref{drude}) arises for $\omega \ll T$,
where we find a delta function at zero frequency in the real part.
Thus the d.c. conductivity is infinite at this order, arising from the collisionless transport of thermally excited
carriers.\\~\\
\noindent
{\bf Exercise:} Evaluate $K_{00}$ from Eq.~(\ref{Kmn}) at $T>0$. First perform the trace over the Dirac matrices,
and then the summation over the frequency. Subtract from your answer the result of integrating over the
frequency; this subtraction can be compensated by the $T=0$ result in Eq.~(\ref{Kmn}). The remaining
expressions are explicitly convergent in the ultraviolet, and the integration over spatial momenta
can be evaluated. Finally, analytically continue the answer to real frequencies to obtain
Eq.~(\ref{drude}).\\

The relaxational processes associated with Eq.~(\ref{relax}) should lead to collisions between the
thermally excited carriers and broaden the delta function at zero frequency. However, this relaxation does
appear in a direct perturbative analysis of the critical theory in powers of $(3-d)$ or $1/N$. As has been discussed
elsewhere\cite{damless,ssqhe,lars1,lars2,vafek1,vafek2,lars3a}, an infinite order resummation is required, whose simplest realization requires solution
of a quantum Boltzmann equation. Such a solution shows that the delta function acquires a width of 
order $(3-d)^2 T$ or $T/N$, and so there is a large d.c. conductivity of order $(3-d)^{-2}$ or $N$.
Thus $\sigma (\omega) $ has the form of `Drude peak' at zero frequency, 
and the behavior in Eq.~(\ref{univ2}) for $\omega \gg T$. However, the accuracy of such a Boltzmann
equation computation is untested, and it is likely that such perturbative analyses of quantum-critical dynamics
are quantitatively unreliable.

We will be satisfied here by scaling arguments which generalize the 
$T=0$ quantum-critical results in Eq.~(\ref{univ2}) to the $T>0$ quantum critical region in Fig.~\ref{qccross}.
The quantum-critical relaxational processes invalidate the form in Eq.~(\ref{univ2}) for the density
correlation function, and we instead expect the form dictated by the hydrodynamic diffusion of charge.
Thus for $K_{00}$, Eq.~(\ref{univ2}) applies only for $\omega \gg T$, while
\beq
K_{00} (\omega, k)  = \chi \frac{D k^2}{Dk^2 - i \omega} \quad, \quad \omega \ll T. \label{diff}
\eeq
Here $\chi$ is the charge susceptibility (here it is the compressibility),
and $D$ is the charge diffusion constant. Associated with Eq.~(\ref{relax}), these have universal
values in the quantum critical region:
\beq
\chi = \mathcal{C}_\chi T \quad , \quad D = \frac{\mathcal{C}_D}{T},
\eeq
where again $\mathcal{C}_\chi$ and $\mathcal{C}_D$ are universal numbers. 
For the conductivity, we expect a crossover from the collisionless critical dynamics at
frequencies $\omega \gg T$, to a hydrodynamic collision-dominated form for $\omega \ll T$.
This entire crossover is universal, and is described by a universal crossover function
\beq 
\sigma (\omega)  = \mathcal{K}_\sigma (\omega/T). \label{Kcross}
\eeq
The result in Eq.~(\ref{univ2}) applies for $\omega \gg T$, and so 
\beq
\mathcal{K}_\sigma (\infty) = \mathcal{K} . \label{Kinf} \eeq
For the hydrodynamic transport, we apply the Kubo formula in Eq.~(\ref{kubo}) to Eq.~(\ref{diff})
and obtain
\beq
\mathcal{K}_\sigma (0) = \mathcal{C}_\chi \mathcal{C}_D
\eeq
which is a version of Einstein's relation for Brownian motion.

\subsection{The AdS/CFT correspondence}
\label{sec:adscft}

Portions of this section have been adapted from Chapter 15 of Ref.~\onlinecite{ssbook2}.

It turns out the AdS/CFT correspondence is `just what the doctor ordered' to compute
strongly-coupled quantum critical dynamics and transport in the orange-colored region
of Fig.~\ref{qccross}. This is a consequence of a crucial property: even at the level of the classical
gravity approximation in the AdS theory, the system relaxes back to thermal equilibrium
in a time which obeys Eq.~(\ref{relax}). No other method in condensed matter physics shares
this remarkable feature. We will review specific computations by this method of the
universal function $\mathcal{K}_\sigma (\omega/T)$, and of the collisionless-to -hydrodynamic 
crossover in the density correlation function.

The CFT solvable by the AdS/CFT correspondence may be viewed as a generalization
of the CFT described by Eq.~(\ref{GN}). It has a closer resemblance to the SU(2) gauge
theory we consider later in Eq.~(\ref{lsu2}).
We take the structure of critical matter fields coupled to a gauge field, and generalize it to a relativistically
invariant model with a non-Abelian SU($N$) gauge group and the maximal
possible supersymmetry. The resulting
supersymmetric Yang-Mills (SYM) theory has only one independent 
coupling constant $g$. Under the RG, it is believed
that $g$ flows to an attractive fixed point at a non-zero coupling $g=g^\ast$; the fixed point then defines
a supersymmetric conformal field theory in 2+1 dimensions (a SCFT3). We are interested here in computing
the transport properties of the SCFT, as a paradigm of quantum critical transport at a strongly interacting 
quantum critical point.

The solution proceeds by a dual formulation as a four-dimensional supergravity theory on a spacetime with uniform negative 
curvature: anti-de Sitter space, or AdS$_4$. The solution is also easily extended to non-zero temperatures,
and allows direct computation of the correlators of conserved charges in real time. 
At $T>0$ a black hole appears in the gravity theory, resulting
in an AdS-Schwarzschild spacetime, and $T$ is the Hawking temperature of the black hole; the real time solutions
also extend to $T>0$.

\begin{figure}
\centering
 \includegraphics[width=4.2in]{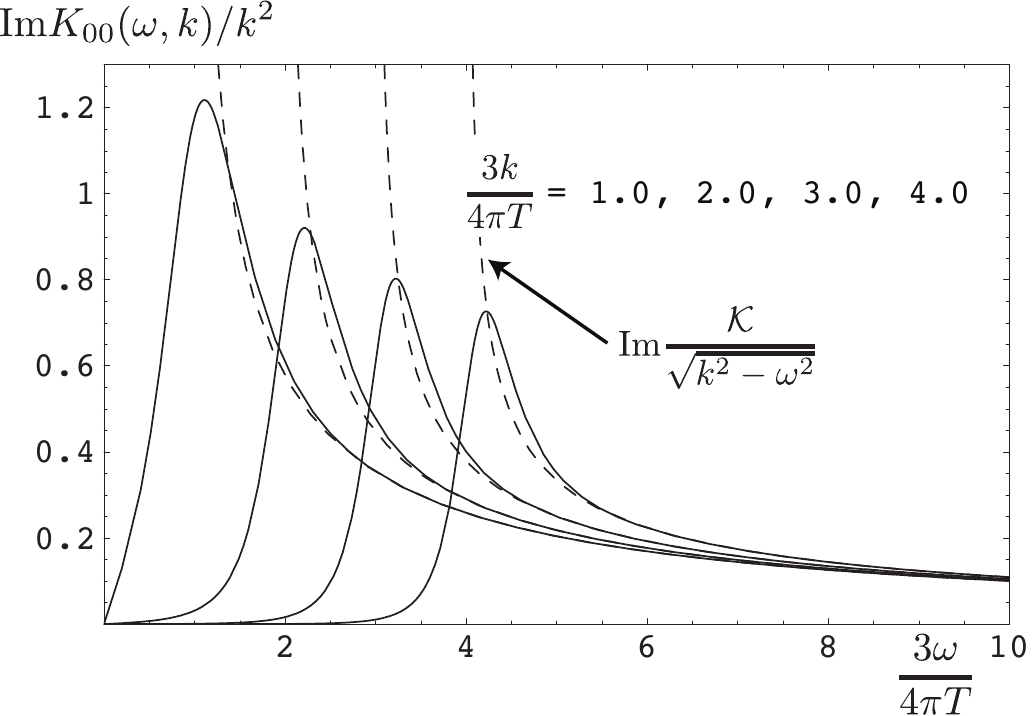}
 \caption{Spectral weight of the density correlation function of the SCFT3 with $\mathcal{N}=8$
 supersymmetry
 in the collisionless regime}
\label{fig:chi_collisionless}
\end{figure}

\begin{figure}
\centering
 \includegraphics[width=4.2in]{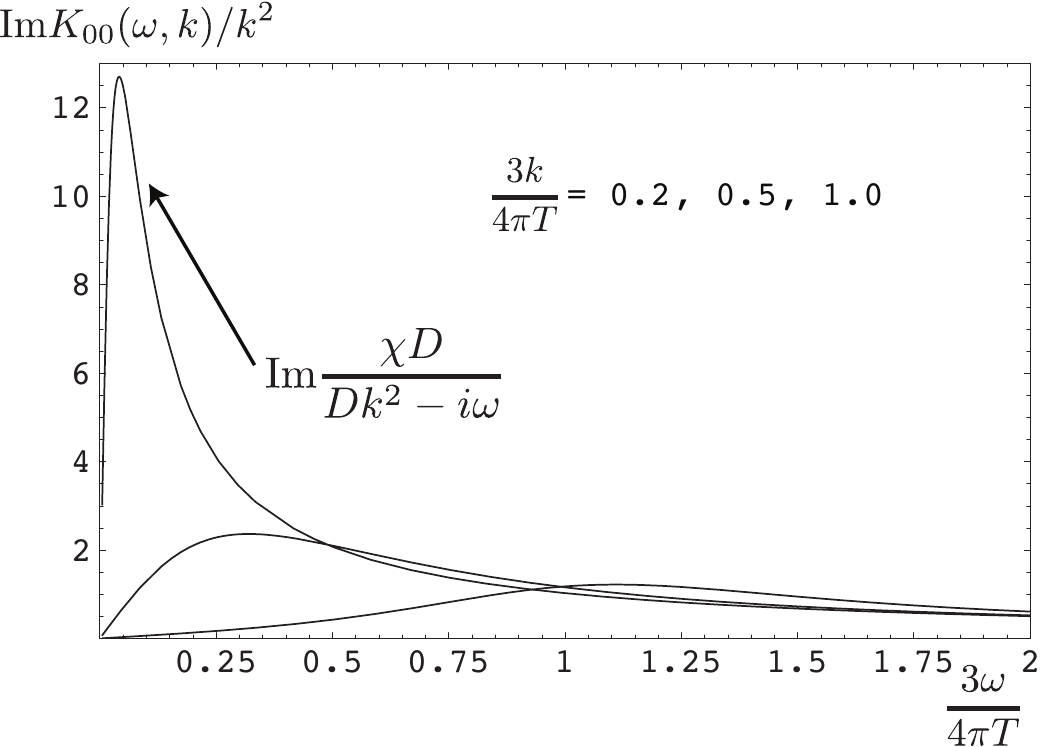}
 \caption{As in Fig.~\ref{fig:chi_collisionless}, but for the collision-dominated regime.}
\label{fig:chi_diff}
\end{figure}

The reader
is referred to the original paper\cite{m2cft}, and to the TASI lectures by Son for an explicit
description of the method. 
In the AdS/CFT correspondence, every
globally conserved quantity in the CFT gets mapped onto a gauge field in AdS. Moreover, in the leading classical
gravity theory on AdS, different global charges commute with each other, and so can be considered separately.
In the end, we have a U(1) gauge field on AdS for every global conservation law of the CFT. The low energy
effective field theory on AdS$_4$ has the standard Einstein-Maxwell action for gravity+electromagnetism:
\begin{equation}
\mathcal{S}_{M} =\frac{1}{g^2_4} \int d^4x \sqrt{-g}\left[-\frac{1}{4}F_{ab}F^{ab} \right]\,.
\label{SEM}
\end{equation}
Here $g_{ab}$ is the AdS-Schwarzschild metric ($g$ is its determinant), $F_{ab}$ is the Maxwell flux tensor,
and $g_4$ is a dimensionless coupling constant fixed by the value of $N$ in the SU($N$) SYM theory.
This 4-dimensional Maxwell theory can be used to compute the 
density correlation function, $K_{00} (\omega,k)$, of the 3-dimensional SYM theory, and the results are shown
in Fig.~\ref{fig:chi_collisionless} and~\ref{fig:chi_diff}.
The most important feature of these results is that the expected limiting forms
in the collisionless (Eq.~(\ref{univ2})) and collision-dominated (Eq.~(\ref{diff})) are obeyed.
Thus the results do display the collisionless to collision-dominated crossover at a frequency of order
$k_B T/\hbar$, as we expected from the physical discussion in Section~\ref{sec:qct}.

At this point, we describe some technical aspects of the results which turn out
to have important physical implications. For this, let us generalize
the constraints on $K_{\mu\nu}$ from current conservation in Eq.~(\ref{ccons}) to 
non-zero temperatures.
At $T>0$, we do not expect $K_{\mu\nu}$ to be relativistically covariant, and so can only constrain it by
spatial isotropy and density conservation.
These two constraints, along with dimensional analyses, lead to the most general form
\begin{equation}
K_{\mu\nu} (\omega, {\bm k} ) =  \sqrt{k^2 - \omega^2} \Bigl( P^T_{\mu\nu}\, K^T (\omega, k)
  + P^L_{\mu\nu}\, K^L (\omega, k) \Bigr),
\label{eq:sscmn}
\end{equation}
where $p_\mu \equiv (-\omega, {\bm k})$ and $k = |{\bm k}|$.
The $K^{L,T}$ are dimensionless functions of the 
arguments, and depend upon $\omega$ and the magnitude of the 2-vector ${\bm k}$.
Also $P^T_{\mu\nu}$ and $P^L_{\mu\nu}$ are
orthogonal projectors defined by
\begin{equation}
P^T_{00} = P^T_{0i} = P^T_{i0}=0~~,~~P^T_{ij} = \delta_{ij} -
\frac{k_i k_j}{k^2}~~,~~P^L_{\mu\nu} =
  \Big(\eta_{\mu\nu} - \frac{k_\mu k_\nu}{p^2}\Big) - P^T_{\mu\nu},
\label{proj}
\end{equation}
with $\eta_{\mu\nu} = \mbox{diag}(-1,1,1)$, and the indices $i,j$ running over the 2 spatial components. Thus, in the general case at $T>0$, the full 
density and current responses are described in terms of 
two functions $K^{L,T} (k, \omega)$, representing current fluctuations longitudinal and transverse to the momentum.
These two functions are not entirely independent. At $T>0$, we expect all correlations to be smooth functions at
$k=0$: this is because all correlations are expected to decay exponentially to zero as a function of spatial separation.
However, this is only possible from (\ref{eq:sscmn}) if we have the additional relation
\begin{equation}
K^T (\omega,0) = K^L (\omega,0).
\label{eq:sslt}
\end{equation}
Finally, we note that application of the Kubo formula in Eq.~(\ref{kubo}) to Eq.~(\ref{eq:sscmn}) yields
\beq
\sigma (\omega) = K^L (\omega, 0).
\label{sigmak}
\eeq

The relations of the previous paragraph are completely general and apply to any theory.
Specializing to the AdS-Schwarzschild solution of SYM3 as determined by the
Einstein-Maxwell theory in Eq.~(\ref{SEM}),
the results were found to obey a simple
and remarkable identity\cite{m2cft}:
\begin{equation}
K^L (\omega,k) K^T (\omega,k) = \mathcal{K}^2
\label{eq:sssdual}
\end{equation}
where $\mathcal{K}$ is a known pure number, independent of $\omega$ and $k$. 
This identity is a consequence of the self-dual structure of Eq.~(\ref{SEM}):
the Maxwell action is in 3+1 dimensions is well-known to have a self-dual structure corresponding
to the exchange of electric and magnetic fields. Thus we have the important and key result that every global charge
in a CFT3 maps onto a self-dual theory in the leading gravity approximation on AdS$_4$. The identity
in (\ref{eq:sssdual}) is a consequence of this emergent self-duality of CFT3s.

The combination of (\ref{eq:sssdual}) and (\ref{eq:sslt}) now fully determine the response functions at
zero momenta: $K^L (\omega,0) = K^T (\omega,0) = \mathcal{K}$. 
Computing the conductivity from Eq.~(\ref{sigmak}), we then have
\beq 
\sigma (\omega ) = \mathcal{K}_\sigma (\omega/ T) = \mathcal{K};
\eeq
{\em i.e.\/} the scaling function in Eq.~(\ref{Kcross}) is {\em independent\/} 
of $\omega$ and equal to the value in Eq.~(\ref{Kinf}).
This result is an important surprise: the conductivity of the classical gravity theory on AdS$_4$ is
frequency-independent. Furthermore, its value is fixed by self-duality to be the constant $\mathcal{K}$
appearing in the self-duality relation (\ref{eq:sssdual}). All these remarkable results are a direct consequence
of the self-duality of the U(1) Maxwell theory on AdS$_4$.

\begin{figure}
\centering
 \includegraphics[width=4.2in]{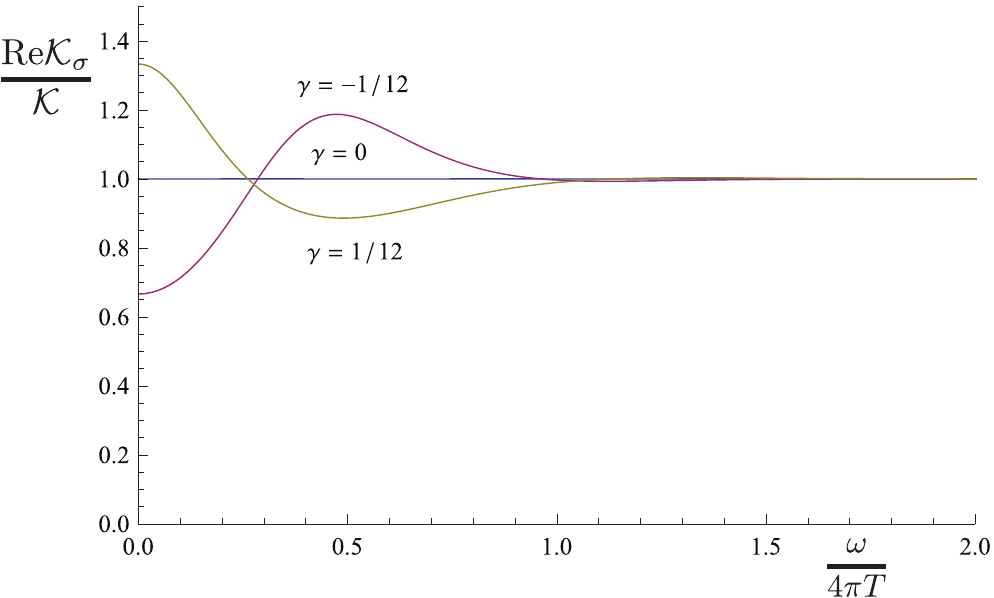}
 \caption{Frequency dependent conductivity for CFT3s for which the AdS$_4$ theory includes the leading
 correction beyond the Einstein-Maxwell theory from Ref.~\onlinecite{myers}. The co-efficient of this correction 
 in the action is $\gamma$, and stability
 requirements impose the bound $|\gamma| < 1/12$.}
\label{fig:myers}
\end{figure}

Given the strong consequences of self-duality relation in Eq.~(\ref{eq:sssdual}), it is useful
to ask whether it can be valid for CFTs beyond those described by the classical Einstein-Maxwell
theory on AdS$_4$. This question was addressed recently by Myers {\em et al.\/}\cite{myers}.
The examined the general structure of the higher-derivative corrections to Eq.~(\ref{SEM}),
and argued that for the current correlations the leading terms could always be transformed into
the following form
which has only {\em one\/} dimensionless constant $\gamma$
($L$ is the radius of AdS$_4$):
\begin{equation}
\mathcal{S}_{M}'=\frac{1}{g^2_4} \int d^4x \sqrt{-g}\left[-\frac{1}{4}F_{ab}F^{ab} +
 \gamma\, L^2  C_{abcd} F^{ab}F^{cd} \right]\,,
\label{eqn4}
\end{equation}
where the extra four-derivative interaction is expressed in terms
of the Weyl tensor $C_{abcd}$. A crucual observation of Ref.~\onlinecite{myers} was that stability and causality constraints on the effective theory
restrict $|\gamma| < 1/12$.
A generalized duality relation applies also to $\mathcal{S}_{M}'$.
However this is {\em not\/} a {\em self\/}-duality.
The dual CFT has current
correlation functions which were characterized by functions $\widetilde{K}^{L,T} (\omega, k)$ which
are distinct from those of the direct CFT $K^{L,T} (\omega, k)$, and the self-duality relation of
Eq.~(\ref{eq:sssdual}) take the less restrictive form
\begin{equation}
K^L (\omega,k) \widetilde{K}^T (\omega,k) = \mathcal{K}^2 \quad , \quad K^T (\omega,k) \widetilde{K}^L (\omega,k) = \mathcal{K}^2.
\label{eq:sssdual2}
\end{equation}
These duality relation determines the correlators of the dual CFT in terms of the direct CFT, but do not fix the latter.
Determination of the functions $K^{L} (\omega, 0) = K^T (\omega,  0) $ requires explicit computation using the
extended theory $\mathcal{S}_{vec}$, and the results for the conductivity are presented in Fig.~\ref{fig:myers}.
Now the conductivity does have a non-trivial universal dependence on $\omega/T$. However, as is clear from
Fig,~\ref{fig:myers}, stability conditions on the effective theory on AdS$_4$ allow only a limited
range of dependence on $\omega/T$. The smooth $\omega/T$ dependence 
in Fig.~\ref{fig:myers} should be contrasted to the
very singular dependence in the free-field result in Eq.~(\ref{drude}); the former is clearly more generic
for a strongly-coupled CFT. It is also interesting to note that the $\omega/T$ dependence in Fig.~\ref{fig:myers}
for $\gamma > 0$ is very similar to the structure we discussed in Section~\ref{sec:qct} on the effect of collisional broadening of 
the singularities in Eq.~(\ref{drude}): the AdS$_4$ result shows a collision-dominated Drude peak at $\omega=0$,
and a collisionless critical continuum at large $\omega$.

Does the duality mapping of Myers {\em et al.}\cite{myers} have an interpretation directly in the CFT,
without using the mapping to AdS$_4$? It has been argued\cite{m2cft,myers,xiyin} that this
duality is the analog of the `particle-vortex' duality of condensed matter physics.
The latter is an exact duality of the critical theory of a complex relativistic scalar field \cite{dasgupta}
(this the theory in Eq.~(\ref{GN}) without the fermions, and with $\varphi^a$ having two components).
In the particle-vortex duality, the world line of the complex scalars is reinterpreted as the world line of 
vortices in the dual theory of a dual complex scalar interacting with an emergent electromagnetic field.
This particle-vortex duality also allows us to interpret the structure of the results in Fig.~\ref{fig:myers}
for $\gamma < 0$. Note that from Eq.~(\ref{eq:sssdual2}) the conductivity of the direct CFT maps onto the 
resistivity of the dual CFT. Thus for $\gamma < 0$, it is the dual CFT which will have a conductivity
which has the structure discussed in Section~\ref{sec:qct}, with a collisionally-broadened Drude peak at $\omega = 0$.
Thus we conclude that a Boltzmann-like picture of transport applies better to the particle interpretation
of the CFT for $\gamma > 0$, and to the vortex interpretation for $\gamma < 0$.

Let us summarize the lessons we have learnt from the AdS theory of quantum critical transport
in strongly interacting systems in 2+1 dimensions. This theory should be view as complementary
to the quasiparticle-based theory, whose implications were discussed in Section~\ref{sec:qct}.
The lessons are: 
\begin{itemize}
\item
There are a large class of strongly interacting 2+1 dimensional quantum liquids which relax back 
to thermal equilibrium in the shortest possible time of order $\hbar /(k_B T)$, as we indicated in (\ref{relax}). 
\item The quasiparticle transport theory\cite{damless,ssqhe,lars1,lars2,vafek1,vafek2,lars3a} starts from the free
theory with an infinite thermal equilibration time, and includes the effect of weak interactions using the Boltzmann
equation. Complementary to this is the quantum-critical transport theory applicable for the shortest possible equilibration
time of order $\hbar/(k_B T)$, which is the classical Einstein-Maxwell theory on AdS$_4$. 
\item The Einstein-Maxwell theory exhibits collisionless dynamics for $\omega \gg T$, and 
collision-dominated dynamics for $\omega \ll T$, as we displayed in Figs.~\ref{fig:chi_collisionless}
and~\ref{fig:chi_diff}.
\item All continuous global symmetries are represented by a self-dual Einstein-Maxwell theory.
\item This emergent self-duality implies that, in systems with particle-hole symmetry, 
$\sigma (\omega)$ is frequency-independent in the Einstein-Maxwell theory
and equal to the self-dual value.
\item For systems with particle-hole symmetry, a frequency dependent conductivity is obtained\cite{myers} upon considering corrections
to the effective Einstein-Maxwell theory, with the forms in Fig.~\ref{fig:myers}. Stability conditions
on the effective theory allow only a limited range of frequency dependence, which depend upon the single
parameter $|\gamma| < 1/12$. For $\gamma > 0$, the frequency dependence has the form expected from
collision-dominated transport of particles, while for $\gamma < 0$ it is characteristic of the transport of vortices. 
It is remarkable that the physical pictures expected from the Boltzmann transport analysis correspond to
precisely those obtained from the holographic approach.
\item Such quantum-critical fluids also have universal momentum transport. By extending the scaling arguments
to momentum transport we would conclude that the ratio of the shear viscosity to the entropy density $\eta/s$
should equal a universal number characterizing the collision-dominated regime. This number was computed
in the Einstein-Maxwell theory by Kovtun {\em et al.}\cite{kss,lars3} and found to equal $\hbar /(4 \pi k_B)$.
The shortest possible relaxation time implies that $\eta$ is also the smallest possible and so these
fluids are `nearly perfect'.
\end{itemize}

\section{U(1) gauge theory and the valence bond solid on the honeycomb lattice}
\label{sec:vbs}

We now return to the honeycomb lattice at half-filling. In Section~\ref{sec:qpt} we described a quantum phase transition
in which two characteristics of the ground state changed simultaneously. In the charge sector, the one electron excitation gap opened
up leading to a transition from the semi-metal to the insulator. And in the spin sector, the breaking of SU(2) spin rotation symmetry
led to N\'eel order in the insulator. However, in the Mott picture, the insulating behavior is tied to repulsion between the 
electrons, which keeps them apart, rather than to any specific symmetry breaking. This would suggest that it is possible
to have an insulating state while preserving spin rotation invariance. We will explore such a possibility in the present section.

Readers not interested in issues related to electron fractionalization and emergent gauge fields
in insulators may skip ahead to the discussion of metallic phases in Section~\ref{sec:tri}.

Our approach will be begin in the N\'eel-ordered insulator, and restore spin rotation invariance by allowing
for slow angular fluctuations in the local {\em orientation\/} of the N\'eel order parameter $\varphi^a$. 
At the same time, it will also pay to transform the fermions to a `rotating reference frame' so that their spin is
measured relative to the orientation of the local N\'eel order\cite{ss,schulz,fsdw}. This transformation is most conveniently done using spinor
variables. So let us decompose the vector N\'eel order $\varphi^a$ into a complex two-component 
bosonic spinor $z_\alpha$ by
\beq \varphi^a = z_\alpha^\ast \sigma^a_{\alpha\beta} z_\beta \label{phiz} \eeq
Such a decomposition is familiar from early work of D'Adda {\em et al.}\cite{dadda}
and Witten\cite{witten}, who established the equivalence between the O(3) non-linear $\sigma-$model
and the CP$^1$ model in 2 spacetime dimensions. A similar equivalence does not immediately
apply in the 3 spacetime dimensional case of interest here because point defects in spacetime
have to be treated with some care\cite{rsb}. In particular, the theory for the fluctuations of the vector field
$\varphi^a$ must allow for point spacetime defects (`instantons') where $\varphi^a = 0$, which are
known in the condensed matter literature as `hedgehogs'. 
Note that these hedgehogs are present even 
in a `fixed-length', non-linear $\sigma$-model in which we set $\varphi^{a2} = 1$;
such models require ultraviolet regularization, and the hedgehogs are invariably permitted in the
regulated theory {\em e.g.\/} with a lattice regularization. 
Ignoring these defects momentarily, let
us proceed as in the earlier work\cite{dadda,witten}. The parameterization in Eq.~(\ref{phiz})
is invariant under the U(1) gauge transformation
\beq z_\alpha \rightarrow z_\alpha e^{i \zeta} \label{zphi} \eeq
and so the theory for the $z_\alpha$ must be a U(1) gauge theory involving an
emergent U(1) gauge field $A_\mu$. The boson only terms in Eq.~(\ref{GN}) are 
equivalent\cite{dadda,witten} to a U(1) gauge theory for the complex scalars $z_\alpha$
\beq \mathcal{L}_z = |(\partial_\mu - i A_\mu) z_\alpha|^2 + s |z_\alpha|^2 + u (|z_\alpha|^2)^2 .
\label{Lz} \eeq
Here the gauge field $A_\mu$ is dynamical, and will acquire a Maxwell action after high energy
$z_\alpha$ modes have been integrated out.

Let us now discuss the point defects. Eqn~(\ref{phiz}) implies\cite{rsb} that the hedgehogs in $\varphi^a$
become Dirac monopoles in $A_\mu$: these are tunnelling events associated with a change in the total
$A_\mu$ flux by $2 \pi$. Such monopoles are permitted by the U(1) gauge theory in Eq.~(\ref{Lz}) only
if the U(1) gauge field is {\em compact\/}. So we must account for the dynamics of such a compact U(1) gauge
theory to completely account for the fluctuations of the local antiferromagnetic order.
The dynamics of the matter fields can suppress the monopoles in some cases\cite{mv,dcp1,dcp2},
and this can then lead to deconfined critical points or phases with a gapless U(1) photon excitation associated with an 
effectively non-compact U(1) gauge field. We will find an example of this phenomenon in Section~\ref{sec:ffl}.

Let us now turn to the fermionic excitations in antiferromagnetic insulator. We transform these
to the rotating reference frame by writing\cite{ss,schulz,fsdw} 
\beq
\left( \begin{array}{c} c_{ \uparrow} \\ c_{ \downarrow}
\end{array} \right) =
 \left( \begin{array}{cc} z_{\uparrow} & -z_{\downarrow}^\ast \\
z_{\downarrow} & z_{\uparrow}^\ast \end{array} \right) \left(
\begin{array}{c} \psi_{ +} \\ \psi_{ -} \end{array} \right)
\label{zpsi}
\eeq
where $\psi_{p}$, $p = \pm$, are the
``electrons'' in the rotating reference frame. The index $p$ measures the spin-projection along the direction of the local
N\'eel order. However, more properly it is the ``charge'' under the emergent U(1) gauge field because
Eq.~(\ref{zpsi}) is invariant under Eq.~(\ref{zphi}) and 
\beq \psi_+ \rightarrow \psi_+ e^{- i \zeta} \quad, \quad \psi_- \rightarrow \psi_- e^{i \zeta}. \label{psiphi} \eeq

An important consequence of Eqs.~(\ref{zpsi}) and (\ref{phiz}) is the identity
\beq
\varphi^a c_{\alpha}^\dagger \sigma^a_{\alpha \beta} c_\beta = (|z_\alpha|^2)^2
\left (\psi^\dagger_+ \psi_+ - \psi^\dagger_- \psi_- \right). \label{rrf}
\eeq
Thus the effective moment acting on the $\psi$ fermions is always along the $z$ axis, as expected by the transformation
to a rotating reference frame.

Let us now take the continuum limit for the fermions in the rotating reference frame. We follow exactly the same
mapping as in Eq.~(\ref{cont}) to map the lattice $\psi$ fermions to continuum 8-component $\Psi$ fermions. 
Based upon Eq.~(\ref{rrf}), we also expect the $\Psi$ fermions to experience a field polarized along the
$\sigma^z$ direction. Combined with gauge invariance, and the structure of Eq.~(\ref{GN}), 
we are led to the following Lagrangian density for
$\Psi$: 
\beq 
\mathcal{L}_\Psi = \overline{\Psi} \gamma_\mu ( \partial_\mu + i \sigma^z A_\mu) \Psi - 
\lambda N_0 \overline{\Psi} \rho^z \sigma^z \Psi. \label{LPsi}
\eeq
Note this is the theory of Dirac fermions of mass $|\lambda N_0|$ coupled to a U(1) gauge field.
The coupling of the fermions to the U(1) gauge field in Eq.~(\ref{LPsi}) can also be derived explicitly by substituting
Eq.~(\ref{zpsi}) into the last term in Eq.~(\ref{GN}), and using the expression for the U(1) gauge field in the CP$^1$ model.
We will present a more explicit derivation of an emergent gauge field for the case of the triangular lattice in Section~\ref{sec:tri} below.

The Lagrangian $\mathcal{L}_z + \mathcal{L}_\Psi$ is then our U(1) gauge theory for the fluctuating N\'eel state,
complementary to the Gross-Neveu-Yukawa theory in Eq.~(\ref{GN}). The remainder of this section is devoted to understanding
the physical properties of $\mathcal{L}_z + \mathcal{L}_\Psi$.

Let us now discuss the phases of this U(1) gauge theory.

First, we have the Higgs phase, where $s<0$ and $z_\alpha$ is condensed. Here the U(1) photon is gapped,
and spin rotation invariance is broken. This is just the insulating N\'eel state, and its properties 
are identical to the N\'eel ordered state described by Eq.~(\ref{GN}).

The other phase with $s>0$ has $z_\alpha$ gapped and spin rotation invariance is preserved.
However, as is clear from Eq.~(\ref{LPsi}), the fermionic spectrum remains gapped.
Thus this phase is clearly not the semi-metal of Eq.~(\ref{GN}). Instead it is a new insulating
phase with spin rotation invariance preserved. Thus we have achieved our objective of describing
an insulator without N\'eel order.

However, this insulator is not a featureless state with a spin and a charge gap, as we will now show.
The interesting physics arises from an interplay of the monopole events with 
the filled band of fermionic states. If we integrate out this filled band via $\mathcal{L}_\Psi$, we generate
an effective Maxwell action for the U(1) gauge field
\beq 
\mathcal{L}_A  = \frac{1}{12 \pi |\lambda N_0|} (\epsilon_{\mu\nu\lambda} \partial_\nu A_\lambda)^2 \label{LA}
\eeq
We recall that the U(1) gauge field is compact, and it was shown by Polyakov\cite{polyakov} that such a gauge
field always acquires a mass gap and is in a confining phase in 2+1 dimensions. 
The confinement is caused by the proliferaction
of monopole tunneling events. Here we will show\cite{rsb,dcp2,cenke1,liang} that the monopole operator has non-trivial transformation
properties under the symmetry group of the honeycomb lattice: consequently, the proliferation of monopoles
in the confining phase breaks a lattice symmetry due to the appearance of valence bond solid (VBS) order.

To see this, it is useful to add an external source $B_\mu$ to the fermion Lagrangian in Eq.~(\ref{LPsi}) so 
that $\mathcal{L}_\Psi$ becomes
\beq
\mathcal{L}_\Psi = \overline{\Psi} \gamma_\mu ( \partial_\mu + i \sigma^z A_\mu) \Psi - 
\lambda N_0 \overline{\Psi} \rho^z \sigma^z \Psi - \frac{i}{2} B_\mu \overline{\Psi} \gamma_\mu \rho^z \Psi . 
\label{LPsi2}
\eeq
This source has been judiciously chosen so that when we integrate out the fermions, the action for the $A_\mu$
gauge field in Eq.~(\ref{LA}) acquires a mutual Chern-Simons term\cite{cenke1,liang}
\beq 
\mathcal{L}_{A}  = \frac{1}{12 \pi |\lambda N_0|} (\epsilon_{\mu\nu\lambda} \partial_\nu A_\lambda)^2 
+ \frac{i}{2 \pi} B_\mu \epsilon_{\mu\nu\lambda} \partial_\nu A_\lambda \label{LB}
\eeq
Let us now proceed with Polyakov's duality mapping on Eq.~(\ref{LB}) to obtain an effective theory of monopoles:
the $B_\mu$ source term will allow us to deduce the connection between the monopole operator and the underlying
lattice fermions. 
The first step
corresponds to decoupling the Maxwell term by a
Hubbard-Stratonovich field, $Y_\mu$, to obtain
\beq
\mathcal{L}_{A}
=  \frac{3|\lambda N_0|}{4 \pi} Y_\mu^2 + \frac{i}{2 \pi} Y_\mu
\epsilon_{\mu\nu\lambda} \partial_\nu A_\lambda   +
+ \frac{i}{2 \pi} B_\mu \epsilon_{\mu\nu\lambda} \partial_\nu A_\lambda \label{LB2}
\eeq
Now, we integrate over $A_\mu$, and this yields the constraint
\begin{equation}
Y_\mu = \partial_\mu \phi - B_\mu.
\label{LB3}
\end{equation}
where $\phi$ is the scalar field which is dual to the photon. We have judiciously chosen
factors of $(2\pi)$ above to ensure a normalization so that $e^{i \phi}$ is the monopole operator.
Finally, inserting Eq.~(\ref{LB3}) into (\ref{LB2}) we obtain\cite{cenke1,liang}
\beq \mathcal{L}_{\phi} =
 \frac{3|\lambda N_0|}{4 \pi} (\partial_\mu \phi - B_\mu)^2. \label{lphoton}
\eeq
In the absence of the external source $B_\mu$ this is a free scalar field theory, which implies that the monopole
operator $e^{i \phi}$ has long-range correlations in 2+1 dimensions. In other words, the free photon phase described by 
Eq.~(\ref{LA}) has a non-zero VEV with $\langle e^{i \phi} \rangle \neq 0$.

\begin{figure}
\center\includegraphics[width=2.2in]{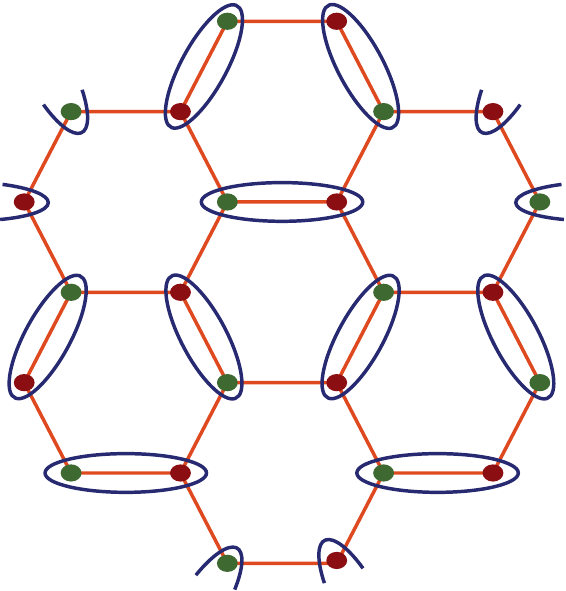}
\caption{A schematic illustration of the valence bond solid (VBS). The ellipses represent singlet valence
bonds between the electrons. These reside preferentially in the pattern shown in the VBS state.
Expectation values of all spin-singlet observables, such as $S^a_i S^a_j$ or 
$c^\dagger_{i \alpha} c_{j \alpha} + c_{j \alpha}^\dagger c_{i \alpha}$, are different on the links with the
ellipses than those without.}
\label{fig:vbs}
\end{figure}

The $B_\mu$ term in Eq.~(\ref{lphoton}) will help us link the monopole operator to the underlying electrons\cite{liang}.
First, we notice that the theory in Eq.~(\ref{LPsi2}) actually enjoys a gauge invariance under which
\bea
\Psi \rightarrow \exp \left( i \frac{ \rho^z}{2} \theta \right) \Psi
\quad , \quad B_\mu \rightarrow B_\mu - \partial_\mu \theta \label{g2}
\eea
where $\theta$ is a field with an arbitrary spacetime dependence. (Note that this gauge invariance
is distinct from that associated with the $A_\mu$ gauge field in Eq.~(\ref{psiphi}), under which $\Psi
\rightarrow \exp( -i \sigma^z \zeta ) \Psi$.) Now we observe that this gauge invariance extends
also to Eq.~(\ref{lphoton}), under which
\beq
e^{i\phi} \rightarrow e^{i \theta} e^{i \phi}. \label{g3}
\eeq
The combination of Eqs.~(\ref{g2}) and (\ref{g3}) now allows us to identify the operator $e^{i \phi}$.
We look for a fermion bilinear of the form $\overline{\Psi} M \Psi$ so it transforms like Eq.~(\ref{g3})
under the gauge transformation in Eq.~(\ref{g2}). Moreover, the Lorentz invariance of the theory
implies that the matrix $M$ should commute with the $\gamma_\mu$ matrices in Eq.~(\ref{defgamma}).
This leads us to the unique choice
\beq
e^{i \phi} \sim \overline{\Psi}  \tau^y ( \rho^x + i \rho^y ) \Psi \sim  \overline{C}  \tau^y ( \rho^x + i \rho^y ) C
\label{mvbs}
\eeq
It now remains to use the geometric definitions in Section~\ref{sec:prelim} and Eq.~(\ref{cont})
to deduce the physical interpretation of the fermion bilinear in Eq.~(\ref{mvbs}). A careful analysis\cite{liang} along these
lines shows that $e^{i \phi}$ is an operator associated with the valence bond solid (VBS) order
in Fig.~\ref{fig:vbs}, and the VEVs of the operators in Eq.~(\ref{mvbs}) imply long-range VBS order.\\~\\
\noindent
{\bf Exercise:} Compute the transformations of the fermion bilinear in Eq.~(\ref{mvbs}) under honeycomb lattice 
symmetries such as translations, reflections, and rotations by 60 degrees. All these transformations map the 
pattern in Fig.~\ref{fig:vbs} either to itself or to 2 equivalents patterns. Assign the weights $1$, $e^{2 \pi i /3}$,
and $e^{4 \pi i /3}$ to these patterns, and show that their transformations coincide with those of Eq.~(\ref{mvbs}).\\

Thus we reach our main conclusion: the insulating phase without N\'eel order as described
by the U(1) gauge theory $\mathcal{L}_z + \mathcal{L}_\Psi$ has long-range VBS order. This order onsets with the
confinement induced by the proliferation of monopoles.

It is interesting to note that the matrices in the fermion bilinears associated with 
VBS order $\sim \overline{C} \tau^y (\rho^x + i \rho^y) C$ and N\'eel order $
\sim \overline{C} \rho^z \sigma^a C$ all anti-commute with each other, and commute
with the $\gamma_\mu$ matrices in Eq.~(\ref{defgamma}). This can be used to formulate the present theory without
gauge fields but using a Wess-Zumino-Witten term\cite{abanov,tanaka,senthilso5,yaolee,liang}: we will not explore this approach here.

We can also use the methods of this section to address the nature of the transition between the N\'eel and VBS
phases. We will not go into details here, but this transition has been proposed\cite{dcp1,dcp2} to be a deconfined critical
point at which monopoles are suppressed, and the critical theory is the non-compact version
of the U(1) gauge theory given by $\mathcal{L}_z$.

\section{SU(2) gauge theory and a phase diagram for the half-filled honeycomb lattice}
\label{sec:su2}

Sections~\ref{sec:afm} and~\ref{sec:vbs} have so far described 3 possible phases of the
honeycomb lattice at half-filling: the semi-metal, the insulator with N\'eel order, and the insulator
with VBS order. The first two phases appear in the theory $\mathcal{L}$ in Eq.~(\ref{GN}),
while the latter two appear in the theory $\mathcal{L}_z + \mathcal{L}_\Psi$ in Eqs.~(\ref{Lz})
and~(\ref{LPsi}). This is an unsatisfactory state of affairs: we would like to write down a single
unified theory in which all 3 phases appear. One approach, implicitly mentioned above, is to extend
Eq.~(\ref{GN}) by including an additional two-component real scalar field representing the VBS
order parameter, and couple it to the fermion bilinear $\overline{C} \tau^y (\rho^x + i \rho^y) C$
appearing in Eq.~(\ref{mvbs}). Integrating out the fermions in the background of a spatially varying 5-component
scalar representing the N\'eel and VBS orders yields a Wess-Zumino-Witten term for the scalar field\cite{abanov,tanaka,senthilso5,yaolee,liang}.
The resulting theory is difficult to work with, and little is known about it in the regime where all three phases can meet.

Here we will present an alternative approach which allows for exotic phases
using an emergent SU(2) gauge field. We will find that the resulting phase diagram
has a fourth semi-metallic phase with an emergent topological order, and an interesting multicritical point.

Our starting point is the observation that the decomposition of the electron into spinful bosons and
spinless charged fermions in Eq.~(\ref{zpsi}) has a larger gauge invariance\cite{fsdw} than U(1).
Rewriting eq.~(\ref{zpsi}) using a natural matrix notation
\beq c = R \, \psi \label{cR}
\eeq
where
\beq R \equiv \left( \begin{array}{cc} z_{\uparrow} & -z_{\downarrow}^\ast \\
z_{\downarrow} & z_{\uparrow}^\ast \end{array} \right), \label{defR}
\eeq
we note that Eq.~(\ref{cR}) is invariant under the gauge transformation generated by
SU(2)$_g$ matrix $U$ under which 
\beq 
R \rightarrow R \, U^\dagger \quad, \quad \psi \rightarrow U \, \psi 
\quad, \quad c \rightarrow c. \label{su2g}
\eeq
This SU(2)$_g$ gauge transformation should be distinguished from the global SU(2) spin rotation $V$, under which
\beq 
R \rightarrow V \, R \quad, \quad \psi \rightarrow \psi \quad , \quad c \rightarrow V \, c. \label{su2}
\eeq

Turning to the N\'eel order $\varphi^a$, this clearly transforms as a ${\bm 3}$ under the global SU(2).
However, the parameterization for the N\'eel order in Eq.~(\ref{phiz}) is not invariant the SU(2)$_g$ gauge 
tranformation in Eq.~(\ref{su2g}). As written, Eq.~(\ref{phiz}) is invariant only under the U(1) gauge transformation
in Eq.~(\ref{zphi}) which was the reason for our original choice of a U(1) gauge theory in Section~\ref{sec:vbs}.
Thus we cannot use Eq.~(\ref{phiz}) as our definition of the N\'eel order in the present SU(2) gauge theory.

It is more natural to proceed here\cite{cenke1,fsdw} by defining the scalar fields using bilinears of the fermions. Thus
using Eq.~(\ref{phiC}) and extending to continuum 8-component fermions near the Fermi points, we define
\beq \varphi^a = \overline{C} \rho^z \sigma^a C \label{phiC2}. \eeq
From this definition it is clear that $\varphi^a$ transforms as a ${\bm 3}$ under the global SU(2),
while it is invariant under SU(2)$_g$, just as expected.

We can also define the corresponding scalar using the $\psi$ fermions\cite{fsdw}:
\beq \Phi^a = \overline{\Psi} \rho^z \sigma^a \Psi \label{PhiPsi}. \eeq
Now $\Phi^a$ transforms as a ${\bm 3}$ under the gauge SU(2)$_g$,
while it is invariant under SU(2).

From Eqs.~(\ref{cR}), (\ref{phiC2}) and (\ref{PhiPsi}), we find that the scalar fields
are related by
\bea
\varphi^a &=& \frac{1}{2} \Phi^b\, \mbox{Tr} \left( \sigma^a R \sigma^b R^\dagger \right) \nn
\Phi^a &=& \frac{1}{2} \varphi^b\, \mbox{Tr} \left( \sigma^b R \sigma^a R^\dagger \right) (|z_\alpha|^2)^2.
\label{phiPhi}
\eea
These relations generalize Eq.~(\ref{phiz}) from the U(1) gauge theory.

To summarize, the matter fields of our SU(2)$_g$ gauge theory are the bosonic matrix $R$, 
the fermions $\Psi$, and the scalar $\Phi^a$. As in Section~\ref{sec:vbs}, we will also
need an emergent dynamic SU(2)$_g$ gauge field $A^a_\mu$.
Using symmetry and gauge invariance, we can now write down the following Lagrangian density\cite{fsdw}
for the SU$_g$ gauge theory; this combines and generalizes the Gross-Neveu-Yukawa model
in Eq.~(\ref{GN}), and the U(1) gauge theory in Eqs.~(\ref{Lz}) and (\ref{LPsi}).
\bea
\mathcal{L}_g &=& \overline{\Psi} \gamma_\mu ( \partial_\mu + i \sigma^a A^a_\mu) \Psi - 
\lambda \Phi^a \overline{\Psi} \rho^z \sigma^a \Psi \nn
&+&  \frac{1}{2} \left[ \left( \partial_\mu \Phi^a - 2 \epsilon_{Abc} A^b_\mu \Phi^c \right)^2 
+ s \, \Phi^{a2} \right]  + \frac{u}{24}  \left( \Phi^{a2} \right)^2 \nn
&+& \mbox{Tr} \left[ (\partial_\mu R - i A_\mu^a R \sigma^a) (\partial_\mu R^\dagger + i A_\mu^a \sigma^a R^\dagger) \right] \nn
&~&~~~~~~~+ \widetilde{s} \, \mbox{Tr} \left( R R^\dagger \right) + \widetilde{u} \, \left[ 
 \mbox{Tr} \left( R R^\dagger \right) \right]^2 . \label{lsu2}
\eea
This Lagrangian combines all three phases discussed so far, and forms the basis of our
remaining discussion of the honeycomb lattice at half-filling.

\begin{figure}
\includegraphics[width=4.5in]{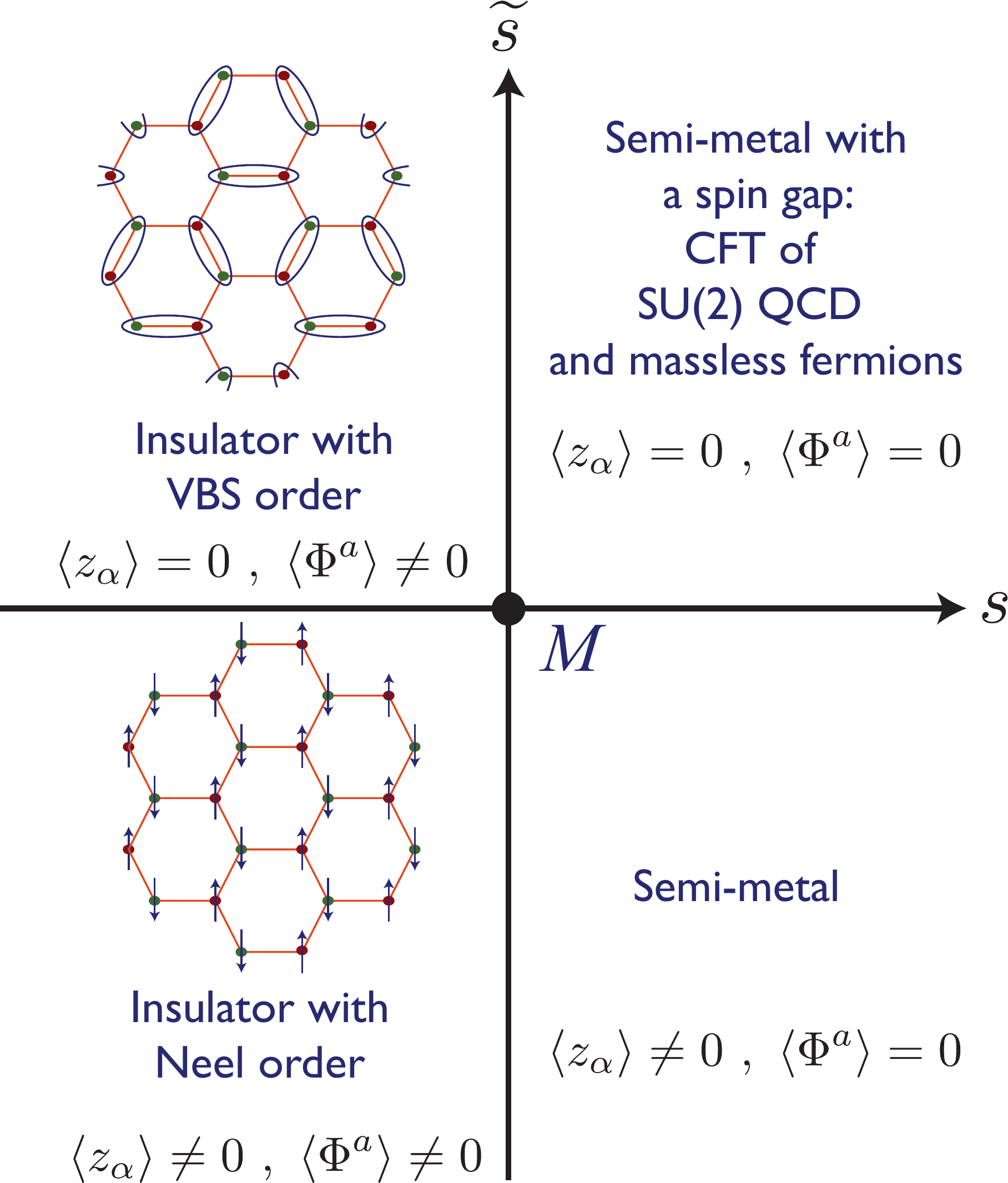}
\caption{Schematic phase diagram of the SU(2)$_g$ gauge theory $\mathcal{L}_g$ in Eq.~(\ref{su2}).
The two phases in the bottom are described by the Gross-Neveu-Yukawa model in Eq.~(\ref{GN}),
while the two phases on the left are described by the U(1) gauge theory $\mathcal{L}_z + \mathcal{L}_\Psi$
in Eqs.~(\ref{Lz}) and (\ref{LPsi}).}
\label{mtheory}
\end{figure}

In mean-field theory, the model $\mathcal{L}_g$ has 4 phases, depending upon whether
one or both of the scalar fields $\Phi^a$ and $R$ are condensed. These 4 phases can be identified
using the methods developed in Section~\ref{sec:qpt} and~\ref{sec:vbs}, and lead to the phase
diagram shown in Fig.~\ref{mtheory}. 
First, we describe how $\mathcal{L}_g$ reproduces the phases
and phase transitions already discussed:
\begin{itemize}
\item The Higgs phase where $\langle R \rangle \neq 0$ breaks SU(2)$_g$ completely. Using SU(2)$_g$ gauge invariance we may as well set $R=1$. Then from Eq.~(\ref{phiPhi}), we have $\Phi^a \sim \varphi^a$, the N\'eel
order parameter. Also the gauge boson $A_\mu^a$ is gapped and can be neglected. Then the theory
$\mathcal{L}_g$ reduces to the Gross-Neveu-Yukawa model in Eq.~(\ref{GN}). As discussed in Section~\ref{sec:qpt},
this theory has semi-metal and insulating N\'eel phases, and these are shown in Fig.~\ref{mtheory}.
\item The Higgs phase where $\langle N^a \rangle \neq 0$ breaks SU(2)$_g$ down to U(1). Then only
the $A^z_\mu$ (say) gauge boson is active, and the theory $\mathcal{L}_g$ reduces to the U(1) gauge
theory $\mathcal{L}_z + \mathcal{L}_\Psi$ discussed in Section~\ref{sec:vbs}. The insulating N\'eel
and insulating VBS phases found there are also shown in Fig.~\ref{mtheory}.
\end{itemize}

The possible new phase of $\mathcal{L}_g$ is the deconfined phase where both $\Phi^a$ and $R$ are
gapped. Then the low energy theory of $\mathcal{L}_g$ is simply massless QCD with the Lagrangian density
\beq \mathcal{L}_{QCD} =  \overline{\Psi} \gamma_\mu ( \partial_\mu + i \sigma^a A^a_\mu) \Psi. \label{QCD}
\eeq
This QCD with a SU(2)$_g$ gauge group with massless 2-component Dirac fermions which carry
$N_c = 2$ colors and $N_f=2$ flavors.
When $N_f$ is large enough, it can be shown from a $1/N_f$ expansion that the confining tendencies of 
non-Abelian gauge fields are screened, and $\mathcal{L}_{QCD}$ describes a non-trivial CFT,
with anomalous dimensions for all observables which are not currents of global flavor or spacetime symmetries. 
It is an open question whether such a critical phase is allowed
for $N_f=2$, as we have assumed in Fig.~\ref{mtheory}. If not, then this phase will be unstable to confinement 
into one of the other phases of Fig.~\ref{mtheory}. If present, this deconfined phase would be a topologically ordered
semi-metal with a spin gap; it is an `algebraic charge liquid' (ACL) in the notation of
Ref.~\onlinecite{rkk2}. The gapless $\Psi$ fermions carry electromagnetic charge, and so there is no
gap to charged excitations excitations and this phase is not an insulator. However, the $\Psi$ fermions are spinless,
and SU(2) spin is only carried by the gapped bosonic excitations; hence the spin gap.

Fig~\ref{mtheory} also shows an interesting multi-critical point $M$, where all 4 phases meet; if the massless QCD phase is confining, this would be the meeting point of 3 phases. Here the SU(2)$_g$ gauge bosons, the scalars
$\Phi^a$ and $R$, and the fermions $\Psi$ are all gapless and critical. Thus $M$ realizes a non-trivial CFT
which can be perturbed by the two relevant directions of the plane of Fig.~\ref{mtheory}. Indeed, it is not unreasonable
to view this multicritical $M$ theory as a non-supersymmetric analog of the M-theory of strings!

\section{The Metal-insulator transition on the triangular lattice}
\label{sec:tri}

This section will describe possible phases of the Hubbard model in Eq.~(\ref{h1}) on the triangular lattice.
We will now consider the case of generic density, so that unlike Sections~\ref{sec:afm},~\ref{sec:vbs}, and~\ref{sec:su2}
on the honeycomb lattice we will allow $\langle n_{i \uparrow} \rangle, \langle n_{i \uparrow} \rangle \neq 1/2$, although
the half-filled density will also appear in our phase diagram.
Unlike the honeycomb lattice, we will ignore the possibilities of magnetically ordered 
phases in which the global SU(2) spin rotation symmetry is broken. The half-filled model on the triangular lattice likely does
have antiferromagnetic order in the limit of large $U$, but we will not consider this complication here. Our purpose here is
to describe the structure of possible phases without magnetic order.

The most significant difference from the honeycomb lattice is apparent in the limit of small $U$, when the electrons are nearly free.
Then the triangular lattice ground state is a metal at all densities, unlike the semi-metal state found on the honeycomb lattice
at half-filling. The semi-metal had a spectrum with a relativistic structure at low energies, a fact which we have exploited in 
our discussion so far. However the metal has zero energy excitations along a line in momentum space, the Fermi surface, and
the fermionic excitations near the Fermi surface do not have a relativistic spectrum. 

Landau's Fermi liquid (FL) theory provides a complete description of the universal properties of the metal. We will 
not review this theory here: the reader is referred to Chapter 18 of Ref.~\onlinecite{ssbook2} for the author's perspective.
A discussion in the context of the gauge-gravity duality appears in a recent paper \cite{liza}. For our purposes here, we need
only two basic facts: ({\em i\/}) the fermionic excitations near the Fermi surface are essentially non-interacting electrons, and
({\em ii\/}) the area enclosed by the Fermi surface is equal to the electron density---this is Luttinger's theorem, which we state more
explicitly below.

\begin{figure}
\center\includegraphics[width=4in]{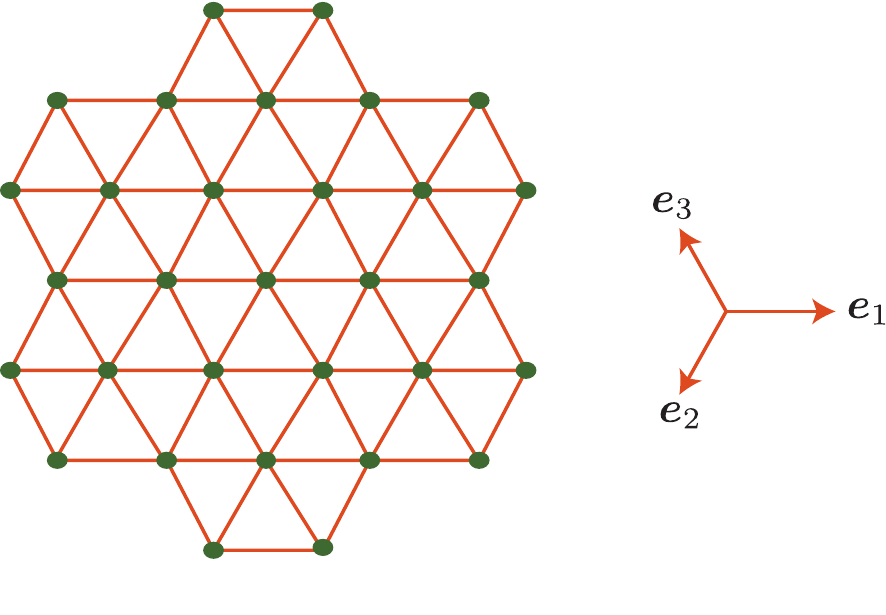}
\caption{The triangular lattice}
\label{fig:trilattice}
\end{figure}

\begin{figure}
\center\includegraphics[width=3in]{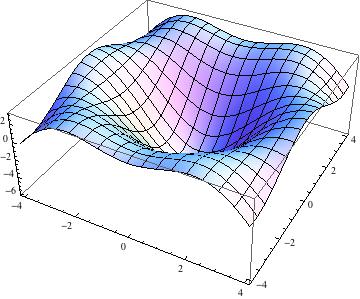}
\caption{The electronic dispersion in Eq.~(\ref{hlattri}) for $\mu=0$ and $t=1$.}
\label{dispt}
\end{figure}

\begin{figure}
\center\includegraphics[width=2.6in]{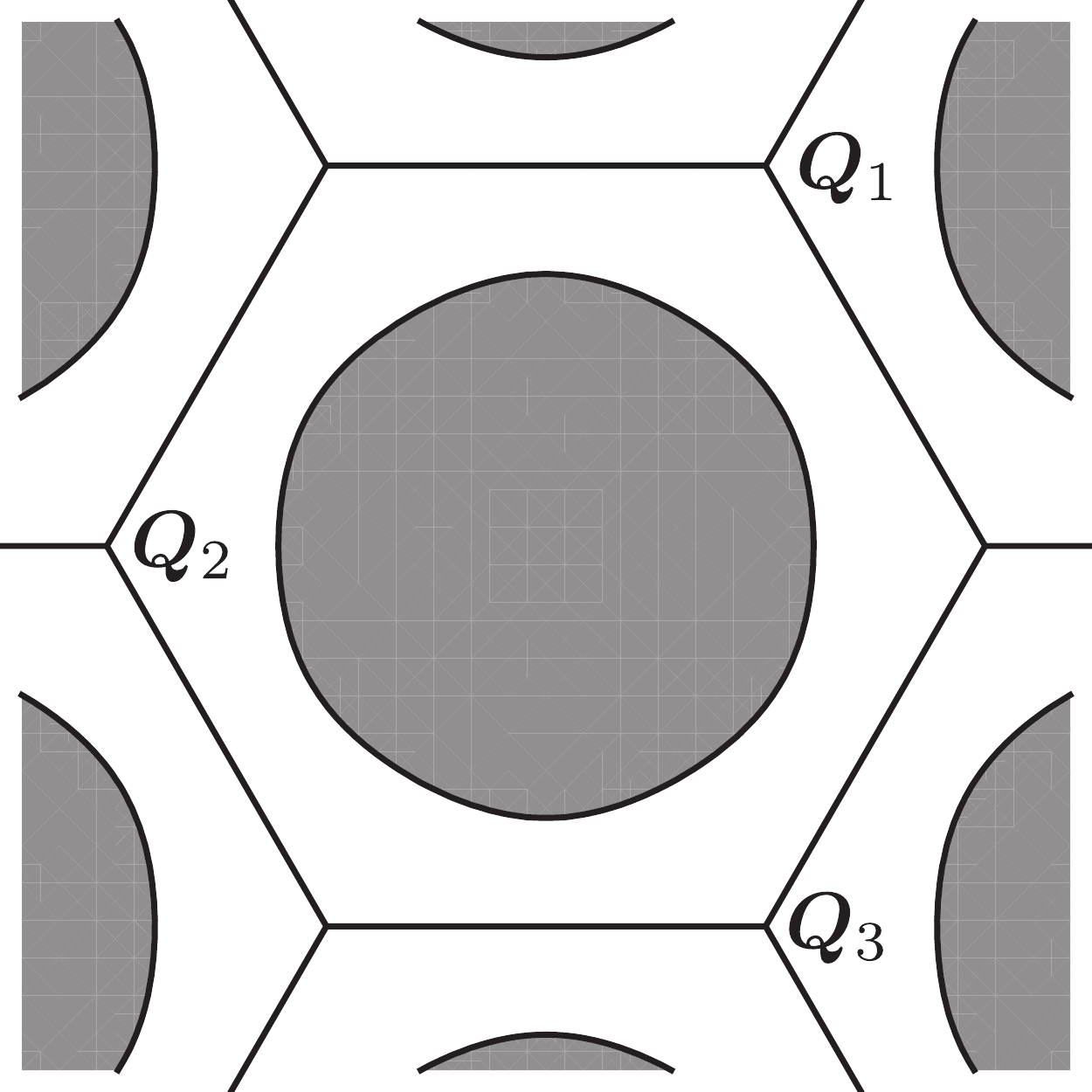}
\caption{The Fermi surface of Eq.~(\ref{hlattri}) for $\mu=1/2$ and $t=1$; the occupied states are shaded. 
Also shown are the periodic images of the Fermi
surface in their respective Brillouin zones.}
\label{fermisurface}
\end{figure}

The FL metal can be described by ignoring the $U$ interactions, and transforming Eq.~(\ref{h1}) to momentum space.
Unlike the honeycomb lattice, there is only one site per unit cell of the triangular lattice, and so the analog of Eq.~(\ref{hlat})
is now simply
\begin{equation}
H_0 = \sum_{{\bm k}} c_\alpha^{\dagger} ({\bm k}) \Bigl[ - \mu  - 2t \Bigl(\cos({\bm k} \cdot {\bm e}_1) + \cos({\bm k} \cdot {\bm e}_2)
+ \cos({\bm k} \cdot {\bm e}_3) \Bigr) \Bigr] c_\alpha ({\bm k} ), \label{hlattri}
\end{equation}
where there are no Pauli matrices associated with sublattice index, the ${\bm e}_i$ are as in Eq.~(\ref{evecs}), and are shown in Fig.~\ref{fig:trilattice}.
The reciprocal lattice now consists of the vectors $\sum_i n_i {\bm G}_i$, where Eq.~(\ref{gvecs}) is replaced by
\begin{equation}
{\bm G}_1 = \frac{4 \pi}{3} ({\bm e}_1 - {\bm e}_2) \quad , \quad 
{\bm G}_2 = \frac{4 \pi}{3} ({\bm e}_2 - {\bm e}_3) \quad , \quad
{\bm G}_3 = \frac{4 \pi}{3} ({\bm e}_3 - {\bm e}_1), \label{gvecstri}
\end{equation}
while the first Brillouin zone is a hexagon with vertices given by Eq.~(\ref{brillQ}), as shown in Fig.~\ref{fermisurface}. 
The electronic dispersion in Eq.~(\ref{hlattri}) is plotted in Fig.~\ref{dispt}: it only has simple parabolic minima at ${\bm k } = 0$,
and its periodic images at ${\bm k} = {\bm G}$, and there are no Dirac points.
At any chemical potential, the negative energy states are occupied, leading to a Fermi surface bounding the set of occupied states, as
shown in Fig.~\ref{fermisurface}.
Luttinger theorem states that the total area of the occupied states, the shaded region of the first Brillouin zone in Fig.~\ref{fermisurface}
occupies an area, $\mathcal{A}$, given by
\begin{equation}
\frac{\mathcal{A}}{2 \pi^2} = \mathcal{N}, \label{landaufl}
\end{equation}
where $\mathcal{N} = \sum_{\alpha} c^\dagger_{\alpha} c_{\alpha}$ is the total electron density. This relationship is obviously true
for free electrons simply by counting occupied states, but it also remains true for interacting electrons, as reviewed recently in 
Ref.~\onlinecite{liza}.

Now we turn up the strength of the interactions, $U$. For the honeycomb lattice, we presented in Sections~\ref{sec:afm2}
and~\ref{sec:qpt} a treatment which allowed for the appearance of spontaneous magnetic moment on each site, leading
to the onset of antiferromagnetic order at large $U$. We also found that the onset of antiferromagnetic order co-incided with 
appearance of the insulator {\em i.e.\/} the semi-metal to insulator transition. This co-incidence was related to the appearance
of a gap in the spectrum even for an infinitesimal antiferromagnetic moment, as in Eq.~(\ref{massivedirac}). 
We can apply a similar treatment here to the triangular lattice. However, such an analysis finds that the onset of magnetic order
does not co-incide with the metal-insulator transition. Instead, we find an intermediate metallic phase with magnetic order,
in which the Fermi surface has been reconstructed into small ``pockets''. Such a reconstruction is observed in many correlated
electron systems. However, we will not explore this route to the insulator here, and refer the reader to 
recent papers \cite{metsdw,fermiology}.

Instead, we will explore here a distinct route to the destruction of the Fermi liquid, one which reaches directly to an insulator
which is a `spin liquid' \cite{mot,lee,senthilmott}. 
The spin liquid insulator is a phase in which the spin rotation symmetry is preserved, and there is a
gap to all charged excitations. In these respects, the spin liquid is similar to the insulating state discussed in 
Section~\ref{sec:vbs}. However, in Section~\ref{sec:vbs} we found that such a insulator had an emergent U(1) gauge field, $A_\mu$,
and the proliferation of monopole defects in $A_\mu$ led to a confining phase in which lattice translational symmetry was
broken by the appearance of VBS order. Here we will find that the triangular lattice spin liquid also has an emergent U(1) gauge field,
but the presence of Fermi surfaces in the spinful excitations leads to the suppression of monopole events. Consequently,
we have a deconfined phase with gapless gauge excitations, no lattice translational symmetry breaking, and the spin liquid
character survives. 

The key to the description of the metal insulator transition is exact rewriting of the Hubbard model in Eq.~(\ref{h1})
as a compact U(1) lattice gauge theory. We proceed with a method which parallels that in Section~\ref{sec:vbs}, of transforming to a `rotating reference frame'.
However, instead of using the frame of reference of local antiferromagnetic order, we use a quantum rotor
to keep track of the charge on each lattice site.  Each rotor has a periodic angular co-ordinate 
$\vartheta_i$ with period $2 \pi$; hence the states of the rotors are $e^{i n_{ri} \vartheta_i}$
where $n_{ri}$ is a rotor angular momentum, whose eigenvalues take all positive and negative integer values. 
We will use the state with all $n_{ri} = 0$
to represent the states with one electron each lattice site. The analog of the transformation to a rotating
reference frame in Eq.~(\ref{zpsi}) is now\cite{florens}
\beq 
c_{ \alpha} = e^{-i \vartheta} f_\alpha \label{cf}
\eeq
where we have dropped the implicit site index, and $f_\alpha$ are neutral fermions (`spinons') which keep track
of the orientation of the electron. We can now identify the 4 states on each lattice site in Eq.~(\ref{4state})
with corresponding states of the rotor and spinons:
\bea 
| 0 \rangle \quad & \Leftrightarrow& \quad e^{- i \vartheta} | 0 \rangle \nn
c^\dagger_{\alpha} | 0 \rangle \quad & \Leftrightarrow& \quad f^{\dagger}_\alpha | 0 \rangle \nn
c^\dagger_{\uparrow} c^\dagger_{\downarrow} | 0 \rangle \quad & \Leftrightarrow& \quad e^{ i \vartheta}
f^\dagger_\uparrow f^\dagger_\downarrow | 0 \rangle \label{rotorrep}
\eea
Note that these allowed states obey the constraint
\beq
f_\alpha^\dagger f_\alpha - n_r = 1.
\label{fnc} 
\eeq
Associated with this constraint is the U(1) gauge invariance which is the analog of Eqs.~(\ref{zphi})
and (\ref{psiphi}):
\beq
f_\alpha \rightarrow f_\alpha e^{i \zeta} \quad , \quad \vartheta \rightarrow \vartheta + \zeta. \label{gauget}
\eeq
Just as in Section~\ref{sec:vbs}, there will be an emergent gauge field $A_\mu$ in the effective theory
of this model. The constraints in Eq.~(\ref{fnc}) will be the Gauss law of this theory.

First, let us rewrite the Hubbard model in terms of these new variables.
Our degrees of freedom are the Fermi operators $f_{i \alpha}$ on each lattice site which 
obey the usual canonical fermion anti-commutation relations as in Eq.~(\ref{h3}), and the rotor angle $\vartheta_i$
and angular momentum $n_{ri}$ which obey
\beq
[ \vartheta_i , n_{rj} ] = i \delta_{ij}.
\eeq
The Hubbard Hamiltonian in Eq.~(\ref{h1}) is now exactly equivalent to
\beq
H [f,\vartheta] = - \sum_{i,j} t_{ij} f_{i \alpha}^\dagger f_{j \alpha} e^{i ( \vartheta_i - \vartheta_j)} + \sum_{i} \left( - \mu (n_{ri} + 1) + \frac{U}{2} n_{ri} (n_{ri} + 1) \right), \label{Hfn}
\eeq 
provided our attention is restricted to the set of states which obey the constraint in Eq.~(\ref{fnc}) on every lattice site;
note that the Hamiltonian in Eq.~(\ref{Hfn}) commutes with constraints in (\ref{fnc}), and so these can be consistently imposed.
In Eq.~(\ref{Hfn}) we have used the rotor angular momentum to measure the charge on each site, and so the dependence of the
energy on $\mu$ and $U$ can be expressed in terms of $n_{ri}$ alone. 

We can now implement the commutation relations, the Hamiltonian, and the constraint in a coherent state path integral
\bea
\mathcal{Z} &=& \int \mathcal D f_{i \alpha} (\tau)  \mathcal D f_{i \alpha}^\dagger (\tau) \mathcal{D} \vartheta_i (\tau) 
\mathcal{D} n_{ri} (\tau) \mathcal{D} \lambda_i (\tau) \exp \Biggl( - \int d \tau \, H [f, \vartheta] \nonumber \\ 
&~&~~~~- \int d \tau \sum_i \Biggl[ f^\dagger_{i \alpha} \frac{\partial f_{i \alpha}}{\partial \tau} 
- i n_{ri} \frac{\partial \vartheta_i}{\partial \tau} + i \lambda_i ( f_{i \alpha}^\dagger f_{i \alpha} - n_{ri} - 1) \Biggr]
 \Biggr),
\label{Ztri}
\eea
where $\vartheta_i (\tau)$ takes values on a circle with unit radius, ensuring quantization of eigenvalues of the angular momentum $n_{ri}$ to integer values. The constraint in Eq.~(\ref{fnc}) is implemented using an auxilliary field $\lambda_i (\tau)$ which acts as a Lagrange multiplier. 

A key observation now is that the partition function in Eq.~(\ref{Ztri}) is invariant under a site, $i$, and $\tau$-dependent
U(1) gauge transformation $\zeta_i (\tau)$ where the fields transform as in Eq.~(\ref{gauget}), and $\lambda$ transforms as
\beq
\lambda_i \rightarrow \lambda_i - \frac{\partial \zeta_i}{\partial \tau}. \label{Lgauge}
\eeq
In other words, $\lambda$ transforms like the temporal component of a U(1) gauge field. 

How do we obtain the spatial components of the gauge field? For this, we apply the Hubbard-Stratonovich transformation of Eq.~(\ref{hs})
to the $t_{ij}$ hopping term in Eq.~(\ref{Hfn}). For this, we introduce another auxiliary complex field $Q_{ij} (\tau)$ which lives on the links of the triangular
lattice and replace the hopping term by
\beq \sum_{i,j} \left( \frac{|Q_{ij} (\tau) |^2}{t_{ij}} - Q_{ij} (\tau) f_{i \alpha}^\dagger f_{j \alpha} - Q_{ij}^\ast (\tau) e^{i (\vartheta_i - \vartheta_j)} \right)
\eeq
We now see from Eq.~(\ref{gauget}), that $Q_{ij}$ transforms under the gauge transformation in Eq.~(\ref{gauget}) as
\beq
Q_{ij} \rightarrow Q_{ij} e^{i (\zeta_i - \zeta_j)}. \label{Qgauge}
\eeq
In other words, arg($Q_{ij}$) is the needed spatial component of the compact U(1) gauge field.

So far, we have apparently only succeeded in making our analysis of the Hubbard model in Eq.~(\ref{h1}) more complicated. Instead of the 
functional integral of the single complex fermion $c_{i \alpha}$, we now have a functional integral over the complex fermions $f_{i \alpha}$,
the rotor $\vartheta_i$, and the auxilliary fields $\lambda_i$ and $Q_{ij}$. How can this be helpful? The point, of course, is that the new
variables help us access new phases and critical points which were inaccessible using the electron operators, and these phases have strong correlations
which are far removed from those of weakly interacting electrons. 

The utility of the new representation is predicated on the assumption that the fluctuations in the auxiliary fields $Q_{ij}$ and $\lambda_i$  
are small along certain directions in parameter space. So let us proceed with this assumption, and describe the structure of the phases so obtained. 
We parameterize
\beq Q_{ij} = \overline{Q}_{ij} e^{A_{ij}} \quad, \quad \lambda_i = - i \overline{\lambda} - A_{i\tau}
\eeq
and ignore fluctuations in the complex numbers $\overline{Q}_{ij}$, and the real number $\overline{\lambda}$. With these definitions, it is clear from Eqs.~(\ref{Lgauge}) and (\ref{Qgauge}) that $A_{ij}$ and $A_\tau$ form the spatial and temporal components of a U(1) gauge field, and so must enter into all physical quantities in a gauge invariant manner. 
The values of $\overline{Q}_{ij}$ and $\overline{\lambda}$ are determined by a suitable saddle-point analysis of the partition function,
and ensure that the constraint (\ref{fnc}) is obeyed.
With these assumptions, the partition function separates into separate fermionic and rotor degrees of freedom interacting
via their coupling to a common U(1) gauge field $(A_{i\tau}, A_{ij})$. In the continuum limit, the gauge fields become a conventional
U(1) gauge field $A_\mu = (A_\tau, {\bm A})$. The partition function of the gauge theory is
\bea
\mathcal{Z} &=& \int \mathcal D f_{i \alpha} (\tau)  \mathcal D f_{i \alpha}^\dagger (\tau) \mathcal{D} \vartheta_i (\tau) 
\mathcal{D} n_{ri} (\tau) \mathcal{D} A_{i \tau} (\tau) \mathcal{D} A_{ij} (\tau) \nonumber \\
&~&~~~~~~~~~~~~~~~~~~ \exp \left( - \int d\tau \Bigl[ \mathcal{L}_f + \mathcal{L}_r 
+ i \sum_i A_{i\tau}\Bigr] \right) \nonumber \\
&~&~~~~\mathcal{L}_f =
\sum_i  f^\dagger_{i \alpha} \left( \frac{\partial }{\partial \tau} + \overline{\lambda} - i A_{i \tau} \right)
f_{i \alpha} - \sum_{ij} \overline{Q}_{ij} f_{i \alpha}^\dagger e^{i A_{ij}} f_{j \alpha} \nonumber \\ 
&~&~~~~\mathcal{L}_r = - i \sum_i n_{ri} \left( \frac{\partial \vartheta_i}{\partial \tau} -  A_{i \tau} \right) 
- \sum_{ij} \overline{Q}_{ij}^\ast e^{i (\vartheta_i - \vartheta_j - A_{ij})} \nonumber \\
&~&~~~~~~~~~~~+ \sum_i \left( -\overline{\lambda} \, n_{ri} - \mu (n_{ri} + 1) + \frac{U}{2} n_{ri} (n_{ri} + 1) \right).
\label{rotorfermiongauge}
\eea
Thus we have fermions $f_{i\alpha}$ moving in a band structure which is roughly the same as that of the electrons in 
Eq.~(\ref{hlattri}), rotors which obey a boson Hubbard-like Hamiltonian, and both are minimally coupled to a
compact U(1) gauge field.

We begin by neglecting the gauge fields, and computing the separate phase diagrams of $\mathcal{L}_f$ and $\mathcal{L}_r$.

The fermions are free, and so occupy the negative energy states determined by the chemical potential $\overline{\lambda}$.

\begin{figure}
\centering
 \includegraphics[width=4.2in]{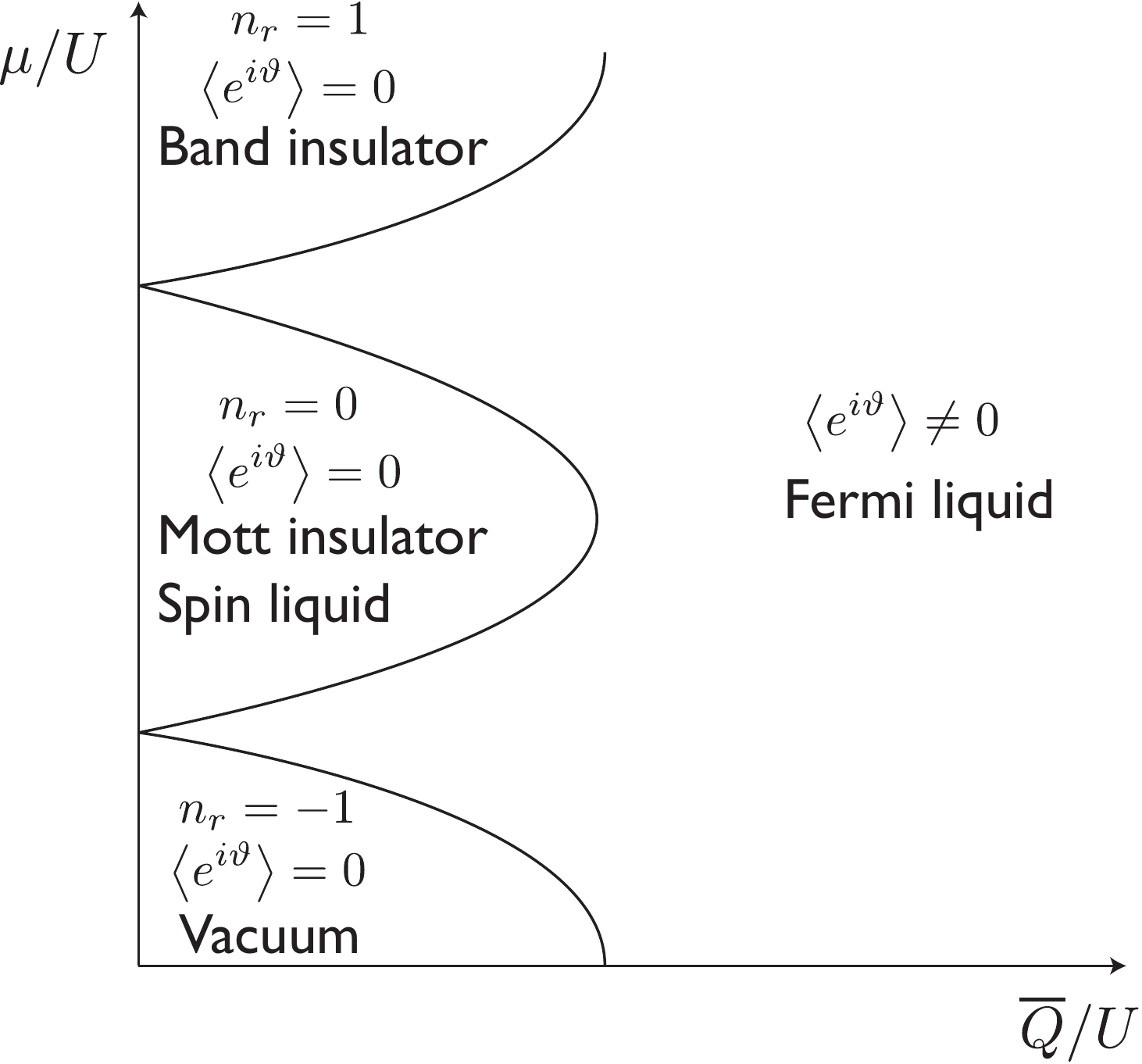}
 \caption{Possible phase diagram of the electron Hubbard model in Eq.~(\ref{h1}) on the triangular lattice.
 This phase diagram is obtained by a mean-field analysis  of the theory $\mathcal{L}_r$ in Eq.~(\ref{rotorfermiongauge}), 
 similar to that for the boson Hubbard model in Chapter 9 of Ref.~\onlinecite{ssbook2}.
 Only the Mott insulating lobes with $n_r = -1,0,1$ are compatible with the constraint in Eq.~(\ref{fnc}); these Mott insulating
 lobes have fermion density $\left\langle f_\alpha^\dagger f_\alpha \right\rangle = n_r + 1$. }
\label{bosemott}
\end{figure}

The phase diagram of $\mathcal{L}_r$ is more interesting: it involves strong interactions between the rotors.
It can be a analyzed in a manner similar to that of the boson Hubbard model (see Chapter 9 of Ref.~\onlinecite{ssbook2}),
leading to the familiar ``Mott lobe'' structure shown in Fig.~\ref{bosemott}.

At large values of $\overline{Q}/U$ we have the analog of the superfluid states of the boson Hubbard model, in which there is a condensate
of the rotor ladder operator $e^{i \vartheta}$.  However, this operator is charged under the U(1) gauge field, and so this phase does
not break any global symmetries. Instead it is a Higgs phase, like the N\'eel phase in the model of Section~\ref{sec:vbs}. In the presence of 
the Higgs condensate, the operator relation in Eq.~(\ref{cf}) implies that $c_\alpha \sim f_\alpha$, and so the $f_\alpha$ fermions carry
the same quantum numbers as the physical electron. Consequently, the $f_\alpha$ Fermi surface is simply an electron Fermi surface.
Furthermore, the Higgs condensate quenches the $A_\mu$ fluctuations, and so there are no singular interactions between the 
Fermi surface excitations. This identifies the present phase as the familiar Fermi liquid, as identified in Fig.~\ref{bosemott}.

Having reproduced a previously known phase of the Hubbard model in the U(1) gauge theory, let us now examine the new phases
within the `Mott lobes' of Fig.~\ref{bosemott}. In these states, the rotor excitations are gapped, and the rotor angular momentum
has integer expectation values. The constraint in Eq.~(\ref{fnc}) implies that only $n_r = -1,0,1$ are acceptable values,
and so only these values are shown. It is clear from the representation in Eq.~(\ref{rotorrep}) that any excitation involving change in electron
number must involve a rotor excitation, and so the rotor gap implies a gap in excitations carrying non-zero electron number. This identifies
the present phases as insulators. Thus the phase boundary out of the lobes in Fig.~\ref{bosemott} is a metal-insulator transition.

Section~\ref{sec:fltrans} will present an explicit demonstration of the insulating and metallic properties of the phases in Fig.~\ref{bosemott}
by a computation of the transport properties of a broader class of models.

The three insulators in Fig.~\ref{bosemott} have very different physical characteristics. 

Using the constraint in Eq.~(\ref{fnc}) we see that
the $n_r =-1$ insulator has no $f_\alpha$ fermions. Consequently this is just the trivial empty state of the Hubbard model,
with no electrons.

Similarly, we see that the $n_r=1$ insulator has 2 $f_\alpha$ fermions on each site. This is the just the fully-filled state of the Hubbard
model, with all electronic states occupied. It is a band insulator.

Finally, we turn to the most interesting insulator with $n_r =0$. Now the electronic states are half-filled, with $\langle f_\alpha^\dagger f_\alpha \rangle = 1$. Thus there is an unpaired fermion on each site, and its spin is free to fluctuate. There is a non-trivial wavefunction in the spin sector,
realizing an insulator which is a `spin liquid'. In our present mean field theory, the spin wavefunction is specified by Fermi surface state
of the $f_\alpha$ fermions. Going beyond mean-field theory, we have to consider the fluctuations of the $A_\mu$ gauge field,
and determine if they destabilize the spin liquid, as we did in 
earlier Section~\ref{sec:vbs}. Here the $f_\alpha$ fermions carry the $A_\mu$ gauge charge, and these
fermions form a Fermi surface. This is a crucial difference from Section~\ref{sec:vbs}, where the
$\psi_\pm$ fermions were gapped. In Section~\ref{sec:vbs} we found that the monopoles proliferated,
leading to confinement and VBS order. Here, the gapless fermionic excitations 
at the Fermi surface prevent the proliferation of monopoles: the low energy fermions suppress
the tunneling event associated with global change in $A_\mu$ flux\cite{hermele,leem}.
Thus the emergent U(1) gauge field remains in a deconfined phase, and this spin liquid state is stable.
These gapless gauge excitations have strong interactions with the $f_\alpha$ fermions, and this
leads to strong critical damping of the fermions at the Fermi surface
which is described by a strongly-coupled field theory\cite{leen,metnem,mross}.
The effect of the gauge fluctuations is also often expressed in terms of an improved trial wavefunction for
the spin liquid \cite{mot}: we take the free fermion state of the $f_\alpha$ fermions, and apply a projection
operator which removes all components which violate the constraint in Eq.~(\ref{fnc}). This yields the 
`Gutzwiller projected' state
\beq
|\mbox{spin liquid} \rangle = \left( \prod_i \left[ \frac{1 - (-1)^{\sum_\alpha f_{i \alpha}^\dagger f_{i \alpha}}}{2} \right] \right)
\left( \prod_{{\bm k} < k_F} f_{\uparrow}^\dagger ({\bm k} ) f_{\downarrow}^\dagger ({\bm k} ) \right) | 0 \rangle,
\eeq
where the product over ${\bm k}$ is over all points inside the Fermi surface. 

Finally, we turn to an interesting quantum phase transition in Fig.~\ref{bosemott}. This is the transition between the spin liquid
and the Fermi liquid at total electron density $\mathcal{N}=1$, which occurs at the tip of the $n_r =0$ Mott lobe. 
From the rotor sector, this looks like a Higgs transition, of the condensation of a complex scalar in the presence of a fluctuating U(1) gauge field. 
However, the fermionic sector is crucial in determining the nature of this transition. Indeed, in the absence of the Fermi surface, this transition would
not even exist beyond mean field theory: this is because the U(1) gauge field is compact, and the scalar carries unit charge, and so the
confining and Higgs phases of this gauge theory are smoothly connected. So we have to combine the Higgs theory of a complex scalar
with the gapless Fermi surface excitations. Introducing a Bose field
\beq
b = e^{-i \vartheta}, \label{btheta}
\eeq
and coarse-graining to a continuum limit for the bosons, we find the field theory \cite{senthilmott}
\bea
\mathcal{L} &=& | ( \partial_\mu + i A_\mu) b|^2 + s |b|^2 + u |b|^4 + i A_\tau \mathcal{N} \nonumber \\
&+& f^\dagger_\alpha \left[ \frac{\partial}{\partial \tau} + \epsilon_f - i A_\tau - \frac{1}{2m_f} 
({\bm \nabla} -  i {\bm A})^2  \right] f_{\alpha}, \label{fermihiggs}
\eea
where the energy $\epsilon_f$ is to be adjusted to yield total fermion density $\mathcal{N} = 1$. 
The transition is accessed by tuning $s$, and we move from a spin liquid for $s < s_c$, to a Fermi liquid
for $s>s_c$. The critical properties of the theory in Eq.~(\ref{fermihiggs}) have been studied \cite{senthilmott,rkk},
and an interesting result is obtained: the Fermi surface excitations damp the gauge bosons so that they become ineffective
in coupling to the critical $b$ fluctuations. Consequently, the gauge bosons can be ignored in the $b$ fluctuations,
and the transition is in the universality class of the 2+1 dimensional XY model.

\section{Fractionalized Fermi liquids}
\label{sec:ffl}

In Section~\ref{sec:tri} we met the canonical description of a compressible metallic state, the Fermi liquid.
The is the state adiabatically connected to the metallic
state of non-interacting electrons. It has long-lived fermionic quasiparticle excitations along the Fermi surface,
and the area enclosed by this Fermi surfaces obeys the Luttinger theorem.

Here we explore an extended model which allows for other compressible phases of strongly interacting electrons at generic densities which
do not break any global symmetries, and which are not adiabatically connected to the limit of non-interacting electrons.
We shall focus here on the compressible state\cite{ffl1,ffl2} known as the fractionalized Fermi liquid (FL*). In principle, the FL* state can appear
in a variety of models of correlated electrons, including ones with a single band, and all the sites equivalent with
$U_i = U$. Such single-band FL* states have been described in recent 
work\cite{rkk1,rkk2,pockets,moon,RW1,RW2,RW3}. However, these single-band
analyses are involved, and require intermediate steps which make them sub-optimal for a first
description of the FL* state. 

\begin{figure}
\centering
 \includegraphics[width=3.8in]{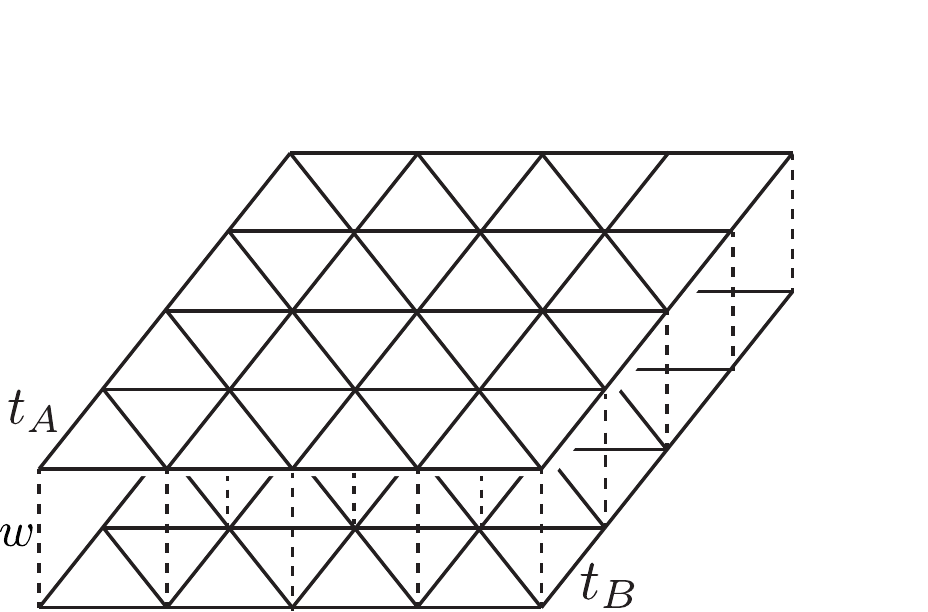}
 \caption{The bilayer triangular lattice. The top layer ($A$) has nearest neighbor hopping $t_A$, the bottom
 layer ($B$) and nearest neighbor hopping $t_B$, and the inter-layer hopping is $w$. A closely related model
 is realized in the experiments of Ref.~\onlinecite{saunders}.}
\label{fig:bilayer}
\end{figure}

Instead, we will introduce the FL* state in a model with 2 types of inequivalent sites. As a simple example,
consider the Hubbard model on a bilayer triangular lattice shown in Fig.~\ref{fig:bilayer}. 
We label the two layers as $A$ and $B$,
and so there are 2 electron operators, $c_{Ai\alpha}$ and $c_{Bi\alpha}$. We write the Hamiltonian 
as
\bea H &=& H_A + H_B + H_{AB} \nn
H_A &=& - t_A \sum_{\langle ij \rangle} c_{Ai\alpha}^\dagger c_{Aj\alpha} + \mbox{H.c.} + 
(\epsilon_A - \mu) \sum_{i} \left( n_{Ai \uparrow} + n_{Ai \downarrow} \right)  \nn
&~&~~~~~~~~~~+
U_A \sum_i \left( n_{Ai\uparrow} - \frac{1}{2}
\right) \left( n_{Ai\downarrow} - \frac{1}{2}
\right) \nn
H_B &=& - t_B \sum_{\langle ij \rangle} c_{Bi\alpha}^\dagger c_{Bj\alpha} + \mbox{H.c.} + 
(\epsilon_B - \mu) \sum_{i} \left( n_{Bi \uparrow} + n_{Bi \downarrow} \right)  \nn
&~&~~~~~~~~~~+
U_B \sum_i \left( n_{Bi\uparrow} - \frac{1}{2}
\right) \left( n_{Bi\downarrow} - \frac{1}{2}
\right) \nn
H_{AB} &=& - w \sum_i  c_{Ai\alpha}^\dagger c_{Bi\alpha} + \mbox{H.c.} \label{hab}
\eea
Here the sites $i,j$ lie on a triangular lattice, and $\langle ij \rangle$ represents the sum over nearest-neighbor pairs.
The Hubbard models on the two layers have distinct values of the hopping parameters, on-site repulsion,
and on-site energies $\epsilon_{A,B}$. Finally, there is an on-site
interlayer tunneling, $w$. Experiments\cite{saunders} on bilayer films of $^3$He adsorbed on graphite provide
a remarkable realization of a closely related model.

First, let us discuss the FL state, where $U_{A,B}$ can be treated perturbatively. Diagonalizing the one-electron Hamiltonian, we find two bands corresponding to the bonding and anti-bonding states between the two 
layers. Let $\mathcal{N}$ be the total density of electrons for each bilayer site of the triangular lattice. So
\beq
\sum_\alpha \left( \left\langle c_{A\alpha}^\dagger c_{A\alpha} \right\rangle +  \left\langle c_{B\alpha}^\dagger c_{B \alpha} \right\rangle \right) = \mathcal{N}. \eeq
This relation holds for every site $i$, and the site-index has been left implicit.
Depending upon the value of $\mathcal{N}$ and interlayer tunneling $w$, 
one or both of the bands will be occupied, leading to one or two Fermi surfaces. Let the areas enclosed
by the Fermi surfaces be $\mathcal{A}_1$ and $\mathcal{A}_2$; if there is only one Fermi surface,
$\mathcal{A}_2 = 0$. Luttinger's theorem fixes the areas of Fermi surfaces to a value which is independent
of the nature of the electron-electron interactions. There is one Luttinger theorem for each global U(1)
symmetry of the Hamiltonian which is not spontaneously broken in the ground state\cite{liza,powell,piers}. Here, the total numbers
of both up-spin and down-spin electrons are separately conserved, and so there are 2 Luttinger constraints.
However, we will implicitly only consider states in which spin rotation invariance is preserved,
and so there is only a single constraint. The constraint is the same as that for non-interacting electrons,
which, as in Eq.~(\ref{landaufl}), is
\beq \frac{\mathcal{A}_1 + \mathcal{A}_2}{2 \pi^2} =  \mathcal{N}. \label{lutt1}
\eeq
We will implicitly assume $\mathcal{N} >1$ below.

We now wish to induce a quantum phase transition to a FL* state. This is most easily done in a model
in which the bottom layer $B$ has a density of one electron per site, while the top layer $A$ remains
dilute (as in the experiment in Ref.~\onlinecite{saunders}). For small hopping this is achieved for
$U_A = U_B = U$ and
\beq
\epsilon_B < \epsilon_A < \epsilon_B + U \,.
\label{eu}
\eeq
Then the bottom layer $B$ will acquire strong electronic correlations like those in Section~\ref{sec:tri}, 
while the dilute gas
on layer $A$ can be treated perturbatively in the two-particle scattering amplitude.
It is customary at this point to follow
the analysis of Section~\ref{sec:afm2}, and project onto this
restricted Hilbert space, while using a canonical transformation to derive an effective Hamiltonian. 
The restricted space has only spin degrees of freedom on $B$ lattice sites,
and as in Section~\ref{sec:afm2}, these spins have exchange interactions with each other. The 
canonical transformation also generates exchange interactions between electrons separate layers,
and this is known as the Kondo exchange interaction. The resulting Hamiltonian is the Kondo-Heisenberg model.
However, we will not take this step here, and continue to work with the Hubbard model 
in Eq.~(\ref{hab}).

We will assume that layer $B$, with a density of one electron per site, realizes the spin liquid state discussed
in Section~\ref{sec:tri}; {\em i.e.\/} it is the $n_r=0$ spin liquid in Fig.~\ref{bosemott}, with a spinon Fermi surface.
We can obtain a description of this spin liquid by applying the analysis of Section~\ref{sec:tri} to layer.
So we replace Eq.~(\ref{cf}) by
\beq 
c_{B \alpha} = e^{-i \vartheta} f_\alpha ,\label{cfb}
\eeq
and perform the same transformations which led to Eq.~(\ref{rotorfermiongauge}).
Then we take the same continuum limit as that used for Eq.~(\ref{fermihiggs}), and obtain the following
continuum Lagrangian $\mathcal{L}$
which captures the low energy physics of the Hubbard model in Eq.~(\ref{hab}).
The degrees of freedom are the $A$ layer electrons $c_{A\alpha}$, the $B$ layer spinons $f_\alpha$,
and the bosonic rotors $b = e^{-i \vartheta}$ as in Eq.~(\ref{btheta}).
The structure of the terms
also follows from general considerations of gauge invariance and the preservation of global symmetries.
\bea
\mathcal{L} &=& \mathcal{L}_f + \mathcal{L}_b + \mathcal{L}_c  + i A_\tau \mathcal{N}_B \nn
\mathcal{L}_f &=& 
f^\dagger_\alpha \left[ \frac{\partial}{\partial \tau} + \epsilon_f - i A_\tau - \frac{1}{2m_f} 
({\bm \nabla}- i {\bm A})^2  \right] f_{\alpha} \nn
\mathcal{L}_b &=&   \Bigl[\left( \partial_\mu - (\epsilon_r - \mu) \delta_{\mu\tau} - i A_\mu + i A_{{\rm ext},\mu}
\right) b^\dagger \Bigr] \nn
&~&~~\times \Bigl[\left( \partial_\mu + (\epsilon_r - \mu) \delta_{\mu\tau} + i A_\mu - i A_{{\rm ext},\mu}
\right) b \Bigr] + 
s |b|^2 + u |b|^4 \nn
\mathcal{L}_c &=& 
c^\dagger_{A\alpha} \left[ \frac{\partial}{\partial \tau} - \mu - i A_{{\rm ext},\tau} - \frac{1}{2m_c} 
({\bm \nabla}- i {\bm A}_{{\rm ext}})^2  \right] c_{A\alpha} \nn
&~&~~~~- w \left( c^\dagger_{A\alpha} b f_\alpha + b^\dagger f_\alpha^\dagger c_{A\alpha} \right) \label{lffl}
\eea
Here $A_\mu = (A_\tau, {\bm A})$ is an emergent U(1) gauge field;
we have also introduced a non-fluctuating
electromagnetic gauge field $A_{{\rm ext},\mu}$ as a source term 
which couples to the current of the globally conserved electromagnetic charge; we have coarse-grained $b$ to a complex scalar field with both amplitude and phase fluctuations;
the symbol $\mu$ refers separately to the chemical potential and spacetime component, and the interpretation
should be clear from the context; the final Yukawa term is allowed by the symmetries, and represents
the inter-layer tunneling $w$; the on-site energies $\epsilon_f$
and $\epsilon_r$ are related to $\epsilon_A$ and $\epsilon_B$ and have to be tuned so that the system
obeys the density constraints to be discussed below.

To review, the continuum theory in Eq.~(\ref{lffl}) has a U(1)$\times$U(1)$_{\rm ext}$ symmetry associated
with the transformations
\bea 
f_\alpha \rightarrow f_\alpha e^{i \zeta} \quad , \quad b &\rightarrow& b^{- i \zeta} \quad , \quad c_{A \alpha} 
\rightarrow c_{A \alpha} \nn
f_\alpha \rightarrow f_\alpha  \quad , \quad b &\rightarrow& b^{ i \tilde\zeta} \quad , \quad c_{A \alpha} 
\rightarrow c_{A \alpha} e^{i \tilde \zeta} \label{gaugec}
\eea
The first U(1) symmetry is gauged by the dynamical emergent U(1) gauge field $A_\mu$, and is the same as that in Eq.~(\ref{gauget}).
The second U(1) symmetry remains global; the fixed external electromagnetic field $A_{{\rm ext},\mu}$
couples a source term which gauges this global symmetry.

In general, there will be 2 Luttinger constraints associated with these two U(1) symmetries\cite{liza,powell,piers}
(as before, we are ignoring spin rotation symmetries here, which is assumed to be always fully preserved).
The first transformation in Eq.~(\ref{gaugec}) leads to a Luttinger constraint on the associated
conserved charge density (which is the continuum
analog of Eq.~(\ref{fnc}))
\beq
\sum_\alpha \left\langle f_\alpha^\dagger f_\alpha \right\rangle - \left\langle \mathcal{Q}_b \right\rangle
= \frac{\mathcal{A}_1}{2 \pi^2} = \mathcal{N}_B . \label{lutt2}
\eeq
Here $\mathcal{N}_B$ is the density of electrons on layer $B$ in the projected Hilbert space: our present
lattice derivation was for $\mathcal{N}_B = 1$, but the continuum theory in Eq.~(\ref{lffl}) is sensible
for any value of $\mathcal{N}_B$. The operator $\mathcal{Q}_b$ is the rotor angular momentum, given by
\beq
\mathcal{Q}_b = - \frac{\partial \mathcal{L}_b}{\partial \mu}.
\eeq
Thus there must be a Fermi surface enclosing area $\mathcal{A}_1$, which counts the density of $f$ fermions
minus the bosonic rotor density.

Similarly, the second transformation of Eq.~(\ref{fnc}) leads to the constraint
\beq
\sum_\alpha \left\langle c_{A\alpha}^\dagger c_{A\alpha} \right\rangle + \left\langle \mathcal{Q}_b \right\rangle
= \frac{\mathcal{A}_2}{2 \pi^2} = \mathcal{N} - \mathcal{N}_B . \label{lutt3}
\eeq
Again, there is a Fermi surface enclosing area $\mathcal{A}_2$ which counts the density of $c_A$ fermions
minus the bosonic rotor density.
Thus our analysis so far appears to imply that there must be at least 2 Fermi surfaces, 
and their areas are constrained by the two independent relations in Eqs.~(\ref{lutt2}) and (\ref{lutt3}).

This last conclusion seems rather surprising from our discussion above of the FL phase.
There we found only a single constraint in Eq.~(\ref{lutt1}) for the total areas of one or more
Fermi surfaces. The only possible conclusion is that the FL phase is {\em not\/} one in which
the U(1)$\times$U(1)$_{\rm ext}$ symmetry of the Lagrangian $\mathcal{L}$ in Eq.~(\ref{lffl})
remains unbroken. Rather the FL phase is realized as a Higgs phase in which the
U(1)$\times$U(1)$_{\rm ext}$ symmetry in Eq.~(\ref{gaugec}) is broken down to a
{\rm diagonal\/} U(1). Just as in Section~\ref{sec:tri}, this is the Higgs phase in which the boson $b$ condenses
\beq
\mbox{$\langle b \rangle \neq 0$ in the FL phase.} \label{bfl}
\eeq
Once the symmetry is broken in this manner, the corresponding Luttinger constraint no longer applies\cite{liza,powell,piers}.
Only the {\em sum\/} of the constraints in Eqs.~(\ref{lutt2}) and (\ref{lutt3}) applies, and this leads immediately
to the defining relation in Eq.~(\ref{lutt1}) of the FL phase. The condensation of $b$ also quenches the emergent
U(1) gauge field, so there are no gapless gauge excitations in the FL state, again as in Section~\ref{sec:tri}.

We now see that state of the theory $\mathcal{L}$ in which the Luttinger constraints in Eqs.~(\ref{lutt2})
and (\ref{lutt3}) apply separately is a new phase: this is the advertised FL* phase, in which the boson $b$
is uncondensed\cite{ffl1,ffl2}
\beq
\mbox{$\langle b \rangle = 0$ in the FL* phase.} \label{bffl}
\eeq
The full U(1)$\times$U(1)$_{\rm ext}$ symmetry is preserved, and the gauge boson $A_\mu$ becomes an
emergent gapless photon. The arguments for the stability of the FL* phase towards gauge fluctuations mirror those
of Section~\ref{sec:tri} for the stability of the spinon Fermi surface in the spin liquid.

The criteria in Eqs.~(\ref{bfl}) and (\ref{bffl}) show that the transition between the FL and FL* states
is tuned by varying the coupling $s$ in $\mathcal{L}_b$ from negative to positive values.
The transition between these phases occurs at a quantum critical point where the scalar
$b$ is also critical.

\subsection{Connections to holographic metals}
\label{sec:holo}

We now connect the above generic theory of the compressible FL and FL* phases of the Hubbard
model to recent studies of compressible metallic phases via the AdS/CFT correspondence.
The discussion below refers to recent work from the gravity perspective; an analysis starting
the canonical supersymmetric gauge theories of gauge-gravity duality may be found in Ref.~\onlinecite{liza}.

A connection was made in Ref.~\onlinecite{ssffl} between a mean-field solution of models like
the Hubbard model in Eq.~(\ref{hab}) and a particular AdS realization of a holographic metal.
Specifically, the bilayer Hubbard model has been solved in a limit with infinite-range
hopping matrix elements between the sites (in contrast to the nearest-neighbor hopping shown
in Eq.~(\ref{hab})). A detailed correspondence was found between the low energy properties
of the FL* phase of such a model and the holographic theory\cite{hong1,faulkner,kachru1} in which the low energy limit
factorized to a AdS$_2 \times$R$^d$ geometry ($d$ is the dimensionality of space). 
This work has been recently reviewed in the companion article\cite{statphys}. 

However, the mean-field solution of Eq.~(\ref{hab}) and the AdS$_2 \times$R$^d$ geometry
share a number of artifacts: they have a non-zero ground state entropy, and the spin correlations
of layer $B$ scale with dynamic exponent $z=\infty$. These artifacts are not expected to be
properties of the field theory $\mathcal{L}$ in Eq.~(\ref{lffl}), applicable for models with short-range interactions.

It is clearly of interest to move beyond the AdS$_2 \times$R$^d$ factorization in the holographic
theory, and derive a holographic model which has a closer correspondence with the phases of the field theory in
Eq.~(\ref{lffl}). A number of recent theories\cite{faulkner,polchinski,gubserrocha,kiritsis,sean1,eric,sean2,kachru2,kachru3,kachru4} 
have examined the feedback of the finite density matter
on the metric of the AdS space, and found that the AdS$_2$ horizon disappears at $T=0$, and is replaced by 
a metric with a finite value of $z$. Many physical properties of such holographic metals
are similar to those of the field theory in Eq.~(\ref{lffl}), but a detailed correspondence awaits future work.

It is useful to consider these recent works in the context of a holographic RG \cite{hong1,faulkner,nickel,karch,heem,mukund}.
In these works, the UV degrees of freedom are coupled to external sources, which in our case is $A_{{\rm ext},\mu}$.
Then an effective action is derived which couples the external sources to the IR degrees of freedom. Two distinct
fixed-point theories have been considered in the literature: those of Nickel and Son\cite{nickel},
and of Faulkner {\em et al.} \cite{hong1} and Faulkner and Polchinski\cite{faulkner}. 
We argue here that these fixed points capture the physics 
of the FL and FL* phases respectively. 

Let us consider, first, the theory of Nickel and Son\cite{nickel}. 
They argued that the low energy theory had an emergent U(1) gauge field,
so that the theory had a U(1)$_{\rm global} \times$U(1)$_{\rm gauge}$ symmetry.
This is strikingly similar to the U(1)$\times$U(1)$_{\rm ext}$ symmetry of Eq.~(\ref{lffl}). 
Indeed, we can more closely map the low energy theory of the FL phase of Eq.~(\ref{lffl})
to the model proposed by Nickel and Son. In the FL phase, we condense the $b$ boson,
and focus on the fluctuations of its phase $b = e^{-i \vartheta}$. Then the effective theory
of the FL phase of Eq.~(\ref{lffl}) is
\beq
\mathcal{L}_{FL} = K_1 \left(\partial_\tau \vartheta - A_\tau + A_{{\rm ext},\tau} \right)^2 
+ K_2 \left({\bm \nabla} \vartheta - {\bm A} + {\bm A}_{{\rm ext}} \right)^2 
+ \Pi_f(A_\mu)  + \mathcal{L}_c \, ,\label{lfl}
\eeq
where $\Pi_f$ is the effective action obtain after integrating out the $f$ spinons.
The structure of Eq.~(\ref{lfl}) is essentially identical
to Eqs.~(6) and (52) of Nickel and Son\cite{nickel}. 

Consider, next, the corresponding low-energy theory of FL* phase. 
Now the $b$ field is not condensed, and has an energy gap, $\Delta$.
So we can safely integrate it out from Eq.~(\ref{lffl}), and obtain an effective theory
for the electrons, the spinons, and the gauge fields:
\bea
\mathcal{L}_{FL*} &=&  \mathcal{L}_f + J_1 \left( c^{\dagger}_{A\alpha} \sigma^a_{\alpha\beta} c_{A\beta} \right)
 \left( f^{\dagger}_{\gamma} \sigma^a_{\gamma\delta} f_\delta \right)
 +  J_2 \left( c^{\dagger}_{A\alpha}  c_{A\alpha} \right)
 \left( f^{\dagger}_{\gamma}  f_\gamma \right) + \mathcal{L}_c \nn
 &+& K_3 \left[ {\bm \nabla} ( A_\tau - A_{{\rm ext}, \tau})
- \partial_\tau ({\bm A} - {\bm A}_{\rm ext} )\right]^2 +
K_4 \left[{\bm \nabla}  \times ({\bm A} - {\bm A}_{\rm ext} )\right]^2 ,
 \label{lfls}
\eea
where $J_1 \sim J_2 \sim w^2/\Delta$. The coupling $J_1$ is the Kondo exchange between the electrons in layer $A$
and the spins on layer $B$, while $J_2$ couples density fluctuations of the two layers. A key property of the FL* 
phase is that the couplings $J_{1,2}$ can be treated perturbatively: there is no flow to strong coupling
in the Kondo exchange, and the layer B spins are not screened by the conduction electrons.
Let us now rewrite the matter component of Eq.~(\ref{lfls}) as
\beq
\mathcal{L}_{FL*} =  \mathcal{L}_f - \frac{1}{2} \left[ F_\alpha^\dagger c_{A\alpha} + c_{A\alpha}^\dagger F_\alpha \right]    + \mathcal{L}_c + \ldots
\label{lfls2}
\eeq
where $F_\alpha$ is a IR fermion defined by\cite{ssffl}
\beq
F_\alpha \equiv - J_1 \left( \sigma^a_{\alpha\beta}  f^{\dagger}_{\gamma} \sigma^a_{\gamma\delta} f_\delta \right) c_{A\beta} - J_2 \left(  f^{\dagger}_{\gamma} f_\gamma \right) c_{A\alpha}.
\label{lfls3}
\eeq
Notice that both fermions in the displayed term in Eq.~(\ref{lfls2}) are invariant under the emergent U(1); this term
is a coupling between the microscopic 
fermion $c_{A\alpha}$ and a composite gauge-invariant fermion operator $F_\alpha$ 
representing the IR
degrees of freedom. We can view $F_\alpha$ in Eq.~(\ref{lfls3}) 
as the most general fermion operator which involves
the IR fermions $f_\alpha$, which is invariant under the gauge transformation associated with $A_\mu$, 
and which also carries the global electron number
charge associated with $A_{{\rm ext},\mu}$.
Then structure of Eq.~(\ref{lfls2})
is precisely that of the semi-holographic theory of Faulkner {\em et al.} \cite{hong1} and Faulkner and Polchinski\cite{faulkner}, and their IR fermion
$F_\alpha$ is chosen by essentially identical criteria.

It would clearly be of interest to also find another fixed point of the holographic theory
corresponding to the quantum-critical points between the FL and FL* phases.

\subsection{Transport theory}
\label{sec:fltrans}

We conclude our discussion of FL and FL* phases by presenting a general formulation of their transport
properties. The arguments below are in the spirit of those of Ioffe and Larkin \cite{ioffe}.

We begin with a theory like $\mathcal{L}$ in Eq.~(\ref{lffl}), and integrate out the matter fields to obtain
a Coleman-Weinberg effective action for the U(1)$\times$U(1) gauge fields $A_\mu$ and $A_{{\rm ext},\mu}$. 
In general, the form of this effective action is constrained only spatial isotropy and gauge invariance. Using the projectors defined in Eq.~(\ref{proj}), we can write the quadratic portion of the effective action in the 
following form (we work here in Euclidean time, and $\omega_n$ is a Matsubara frequency)
\bea
\mathcal{S} &=& \frac{1}{2} T\sum_{\omega_n} \int \frac{d^2 k}{4 \pi^2} \Biggl[
A_\mu \Bigl( P_{\mu\nu}^L K^L_f (\omega_n, k) + P_{\mu\nu}^T K^T_f (\omega_n, k) \Bigr) A_\nu \nn
&~&~~~+ (A_\mu - A_{{\rm ext},\mu} ) \Bigl( P_{\mu\nu}^L K^L_b (\omega_n, k) + P_{\mu\nu}^T K^T_b (\omega_n, k) \Bigr) 
(A_\mu - A_{{\rm ext},\mu} ) \nn
&~&~~~~~~~~~~~~~+ A_{{\rm ext},\mu}  \Bigl( P_{\mu\nu}^L K^L_c (\omega_n, k) + P_{\mu\nu}^T K^T_c (\omega_n, k) \Bigr) 
A_{{\rm ext},\mu}  
\Biggr] \label{saa}
\eea
Here $K_f^{L,T}$ are given by the correlator of the current of the $f$ fermions,
$K_c^{L,T}$ by the correlator of the current of the $c_A$ fermions, and $K_b^{L,T}$ by the current
of the bosonic $b$ rotors. 
Note that, unlike Eq.~(\ref{eq:sscmn}), we have not pulled out a factor of $\sqrt{k^2 + \omega_n^2}$ in the
definition of the $K^{L,T}$.
In general, determining these functions requires a complex transport analysis of the theory
in Eq.~(\ref{lffl}). However, in the FL and FL* phases, the simpler low energy effective theories in Eqs.~(\ref{lfl})
and (\ref{lfls}) lead to simple forms for the bosonic correlators $K_b^{L,T}$.

In the FL phase, integrating out the phase $\vartheta$ in Eq.~(\ref{lfl}) we obtain
\beq
K_b^L (\omega_n, k) = \frac{K_1 K_2 (k^2 + \omega_n^2)}{K_2 k^2 + K_1 \omega_n^2} \quad
, \quad  K_b^T (\omega_n, k) = K_2. \label{Khiggs}
\eeq
Thus $K_b^{L,T}$ are constants in the limits of small momenta or frequency. Indeed, had we chosen the
velocity of `light' judiciously in the definition of $P^L_{\mu\nu}$ in Eq.~(\ref{proj}), we would have had
$K_b^L (\omega_n)  = K_1$.

In contrast, in the FL* phase, we can directly match the low energy theory in Eq.~(\ref{lfls}) to 
Eq.~(\ref{saa}) and obtain
\beq
K_b^L (\omega_n, k) = K_3 (k^2 + \omega_n^2) \quad
, \quad  K_b^T (\omega_n, k) = K_3 \omega_n^2 + K_4 k^2 .
\eeq
Now the $K_b^{L,T}$ vanish in the limit of small momentum and frequency.

We need to use the respective low energy theories of the FL and FL* phases in Eqs.~(\ref{lfl}) and (\ref{lfls})
to determine $K_c^{L,T}$ and $K_f^{L,T}$, and then combine them with the above results
for $K_b^{L,T}$ to obtain the physical conductivity. As in Nickel and Son\cite{nickel},
and in Ioffe and Larkin\cite{ioffe}, this is obtained by implementing the equation of motion
of the emergent gauge field $A_\mu$ in Eq.~(\ref{saa}). 
This equation of motion is equivalent to the constraint that the current of the $b$ bosons
must equal the current of the $f$ fermions, which is a consequence of the lattice constraint
in Eq.~(\ref{fnc}). Evaluating the equation of motion $\delta \mathcal{S}/\delta A_\mu = 0$ from
Eq.~(\ref{saa}), and substituting the resulting value of $A_\mu$ back (after suitable gauge fixing), 
we obtain an effective
action for the probe field $A_{{\rm ext}, \mu}$ alone
\beq
\mathcal{S}_{{\rm ext}} = \frac{1}{2} T\sum_{\omega_n} \int \frac{d^2 k}{4 \pi^2}  A_{{\rm ext},\mu}  \Bigl( P_{\mu\nu}^L K^L_{{\rm ext}} (\omega_n, k) + P_{\mu\nu}^T K^T_{{\rm ext}} (\omega_n, k) \Bigr) 
A_{{\rm ext},\mu}  
\eeq
with 
\bea
K^L_{{\rm ext}} (\omega_n, k) &=& K_c^L (\omega_n,k) + \frac{K_f^L (\omega_n, k) K_b^L (\omega_n, k)}{K_f^L (\omega_n, k) + K_b^L (\omega_n, k)} \nn
K^T_{{\rm ext}} (\omega_n, k) &=& K_c^T (\omega_n,k) + \frac{K_f^T (\omega_n, k) K_b^T (\omega_n, k)}{K_f^T (\omega_n, k) + K_b^T (\omega_n, k)} \label{ilgen}
\eea
After analytic continuation to Minkowski space, 
these results lead to the physical conductivity via the Kubo formula in Eq.~(\ref{kubo})
\beq
\sigma (\omega) = \frac{i}{\omega} K^L_{{\rm ext}} (\omega, 0). \label{sigmaext}
\eeq
The distinction from Eq.~(\ref{sigmak}) is a consequence of omitting in Eq.~(\ref{saa}) the
prefactor $\sqrt{\omega_n^2 + k^2}$ present in Eq.~(\ref{eq:sscmn}).

These expressions can be used along with specific computations of the dynamics 
of the $f$ and $c$ fermions: the latter can be carried out either using a Boltzmann theory of 
the continuum model in Eq.~(\ref{lffl}), or via a theory on AdS. The analysis by Nickel and Son \cite{nickel} 
for their holographic theory is equivalent to the application of Eq.~(\ref{ilgen}). Let us verify that the present method
yields the expected FL behavior in the Higgs phase where Eq.~(\ref{Khiggs}) implies that $K_b^L (\omega, 0) = K_2$.
We assume that $c$ and $f$ Fermi surfaces have metallic conduction with $K_c^L (\omega, 0) = -i \omega \sigma_c$
and $K_f^L (\omega, 0) = -i \omega \sigma_f$, with $\sigma_{c,f}$ the respective conductivities. Inserting these expressions in 
Eq.~(\ref{sigmaext}), we obtain the expected FL behavior with $\sigma = \sigma_c + \sigma_f$ in the limit $\omega \rightarrow 0$.
Thus there is no superfluidity associated with the condensation of $b$, and the gauge fluctuations lead eventually to metallic behavior.
Similarly, it is easy to show that if $b$ and $c$ excitations are gapped, we have insulating transport, even though the
$f$ spinons have a gapless Fermi surface.

Further theoretical work exploring the connection between the AdS and Boltzmann approaches to transport is clearly of interest.

\subsection*{Acknowledgements}

I am very grateful to the participants of TASI 2010 in Boulder, and of the ICTS Chandrasekhar Lecture 
Series and Discussion Meeting on ``Strongly Correlated Systems and AdS/CFT'' in Bangalore, Dec 2010.
Many of 
the ideas presented here were developed and sharpened during discussions at these meetings.
This research was supported by the National Science Foundation under grant DMR-0757145 and by a MURI grant from AFOSR.

\end{document}